\documentclass[12pt,preprint]{aastex}

\begin{document}

\title{MMTF: The Maryland$-$Magellan Tunable Filter}

\author{S.\ Veilleux\altaffilmark{1,2}, B.\ J.\
  Weiner\altaffilmark{1,3}, D.\ S.\ N.\ Rupke\altaffilmark{1,4}, M.\
  McDonald\altaffilmark{1}, C.\ Birk\altaffilmark{5}, J.\
  Bland-Hawthorn\altaffilmark{6,7}, A.\ Dressler\altaffilmark{5}, T.\
  Hare\altaffilmark{5}, D.\ Osip\altaffilmark{8}, C.\
  Pietraszewski\altaffilmark{9}, and S.\ N.\ Vogel\altaffilmark{1}}

\altaffiltext{1}{Department of Astronomy, University of Maryland,
  College Park, MD 20742; veilleux@astro.umd.edu,
  mcdonald@astro.umd.edu, vogel@astro.umd.edu}

\altaffiltext{2}{Also: Max-Planck-Institut f\"ur Extraterrestrische
  Physik, Postfach 1312, D-85741 Garching, Germany}

\altaffiltext{3}{Current Address: Steward Observatory, University of
  Arizona, 933 N. Cherry Ave., Tucson, Arizona 85721;
  bjw@as.arizona.edu}

\altaffiltext{4}{Current Address: Institute for Astronomy, University
  of Hawaii, 2680 Woodlawn Drive, Honolulu, HI 96822;
  drupke@ifa.hawaii.edu}

\altaffiltext{5}{Observatories of the Carnegie Institution for
  Science, 813 Santa Barbara Street, Pasadena, CA 91101;
  birk@obs.carnegiescience.edu, dressler@obs.carnegiescience.edu,
  thare@obs.carnegiescience.edu}

\altaffiltext{6}{School of Physics, University of Sydney, Sydney,
    NSW, Australia; jbh@physics.usyd.edu.au}

\altaffiltext{7}{Anglo-Australian Observatory, P. O. Box 296, Epping,
    NSW 2121, Australia}

  \altaffiltext{8}{Las Campanas Observatory, Carnegie Observatories,
    Casilla 601, La Serena, Chile; dosip@lco.cl}

\altaffiltext{9}{IC Optical Systems Ltd, 190-192 Ravenscroft Road,
    Beckenham, Kent BR3 4TW, United Kingdom;
    chris.pietraszewski@icopticalsystems.com}

\begin{abstract}
  This paper describes the Maryland-Magellan Tunable Filter (MMTF) on
  the Magellan-Baade 6.5-meter telescope.  MMTF is based on a 150-mm
  clear aperture Fabry-Perot (FP) etalon that operates in low orders
  and provides transmission bandpass and central wavelength adjustable
  from $\sim$5 to $\sim$15 \AA\ and from $\sim$5000 to over $\sim$9200
  \AA, respectively. It is installed in the Inamori Magellan Areal
  Camera and Spectrograph (IMACS) and delivers an image quality of
  $\sim$0$\farcs$5 over a field of view of 27\arcmin\ in diameter
  (monochromatic over $\sim$10\arcmin). This versatile and
  easy-to-operate instrument has been used over the past three years
  for a wide variety of projects. This paper first reviews the basic
  principles of FP tunable filters, then provides a detailed
  description of the hardware and software associated with MMTF and
  the techniques developed to observe with this instrument and reduce
  the data. The main lessons learned in the course of the
  commissioning and implementation of MMTF are highlighted next,
  before concluding with a brief outlook on the future of MMTF and of
  similar facilities which are soon coming on line.
\end{abstract}


\section{Introduction}

The science drivers for tunable narrowband imaging of the night sky
are very strong: (1) isolate the faint signal (e.g.\ emission line
signatures) from distant objects against the bright noise from the sky
or the source itself (e.g. continuum emission); (2) achieve high survey
efficiency by combining high throughput, large field of view (FOV) and
good spectral resolution (Jacquinot 1954); (3) produce well-defined
volume-limited samples of objects selected by the wavelength
(redshift) coverage of the instrument and unaffected by the
morphological (compact versus diffuse) and/or broadband brightness
biases generally associated with multi-object spectroscopy. A judicious
choice of emission lines also helps select targets by precisely the
quantity that one tries to measure (e.g.  H$\alpha$, [O~II]
$\lambda\lambda$3727, or Ly$\alpha$ for the star formation rates in
galaxies; Jones \& Bland-Hawthorn 2001).

The ideal tunable filter is an imaging device which can isolate an
arbitrary spectral band $\delta\lambda$ at an arbitrary wavelength
$\lambda$ over a broad, continuous spectral range, preferably with a
response function which is identical in form at all wavelengths. As
described in Bland-Hawthorn (2000a, 2000b), a rich variety of physical
phenomena can isolate a finite spectral band: absorption, scattering,
diffraction, evanescence, birefringence, acousto-optics, single-layer
and multilayer interference, multi-path interferometry, polarizability
and so on. All spectroscopic techniques rely ultimately on the
interference of beams that traverse different optical paths to form a
signal. The technologies which come closest to the ideal tunable
filter are the air-gap Fabry-Perot (FP) and Michelson (Fourier
transform) interferometers.

Recent technological developments in the fabrication of
high-performance FP etalons\footnote{ Alignment between the two highly
  polished plates of glass in these etalons can be maintained down to
  spacings of 2 $\mu$m (measured at the coating surfaces).  Long-range
  stacked piezo-electric transducers (PZT) allow the parallel plates
  to be scanned over a physical spacing of 4 $\mu$m (5 wavelengths or
  10 interference orders in the I band).  Reflective coatings are laid
  down with ionic bombardment which allows for very high integrity and
  uniformity in the coating response over a broad wavelength range.}
have allowed the implementation of reliable and cost-effective FP
tunable filters with wide monochromatic FOV and adjustable
transmission characteristics.  These low-order FP etalons are the most
straightforward application of tunable filter technology.  Other, more
sophisticated techniques such as Lyot and acousto-optic filters exist,
but all are currently considerably more expensive, more risky, and
typically have lower throughput.

Deep optical emission-line surveys with the prototype FP tunable
filter, the Taurus Tunable Filter (or TTF; Bland-Hawthorn \& Jones
1998a, 1998b; Bland-Hawthorn \& Kedziora-Chudczer 2003) on the 3.9m
Anglo-Australian Telescope, have shown that there is a rich field of
science awaiting exploration with large ground-based telescopes
equipped with these filters.  In this paper, we describe the first
such tunable filter in routine operation on a large telescope: the
Maryland-Magellan Tunable Filter (MMTF) on the Magellan Baade
6.5-meter telescope at Las Campanas.  MMTF was commissioned in June
2006 and has been used as a PI instrument since January 2007.  A
number of engineering runs over the past two years have allowed us to
fine tune MMTF and make it easier to use. This versatile
and easy-to-use instrument is currently being used for science at a
rate of $\sim$10 nights per semester. 

In this paper, we summarize the technical aspects of MMTF, discussing
them primarily from a user's point of view. In Section 2, we introduce
the basic concepts behind FP tunable filters. In Sections 3 and 4, we
describe the hardware and software associated with MMTF and the
techniques used to observe with this instrument and reduce the data.
In Section 5, we highlight the main lessons learned in the course of
the commissioning and implementation of MMTF. In Section 6, we
conclude with a discussion of the future of MMTF and of similar
facilities which are coming on line in the near future. The official
website of MMTF is http://www.astro.umd.edu/$\sim$veilleux/mmtf/ and
should be consulted for more detail.

\section{Basic Principles of  FP Tunable Filters}

FP tunable filters operate on similar principles to narrow-band
interference filters. In a single-plate interference filter, the
interference occurs in the interior of a solid plate, where the two
sides of the plate act as reflective surfaces 
(e.g.\ MacLeod 2001). In a FP tunable filter, the interference arises
in the air gap between the surfaces of two separate plates of high
reflectivity. Light travels through the interferometer such that at
each radius from the optical axis a specific wavelength is imaged,
resulting in the wavelength, $\lambda$, varying linearly with the
cosine of the angle from the optical axis, $\theta$ (At the detector, the
$\theta$ dependence translates into a radial dependence of wavelength
via the camera focal length; the optical axis is at the center of this
radial pattern):
\begin{equation}
m\, \lambda = 2\, \mu\, l\, {\rm cos}\, \theta,
\label{eqn1}
\end{equation}
where $m$ is the order number, $l$ is the spacing between the two
plates, and $\mu$ is the index of refraction of the air gap.  To first
approximation, $\mu \approx 1$ and $\theta \approx 0$ so $l \approx m
\lambda/2$. Low-order etalons of tunable filters require small plate
spacings of a few microns.

From eqn.\ (\ref{eqn1}), assuming $\mu \simeq 1$ and $\theta \simeq
0$, one finds that interference orders are separated by a free
spectral range, $FSR = \lambda/m \simeq \lambda^2/2l$.  Each
transmission peak is characterized by an Airy profile. The ratio of
the free spectral range to the instrumental resolution
is the effective finesse, $N$, or the number of independent resolution
elements within one {\em FSR}. In a perfect system, the effective
finesse is equal to the reflective finesse, $N_R$, which is only a
function of the reflection coefficient, $\Re$, of the coatings: $N_r =
\pi \Re^{1/2}/(1-\Re)$. However, irregularities in the plate surfaces
and coating thicknesses, as well as deviations from perfectly parallel
plates (e.g.\ bowing), degrade the effective finesse, and hence the
efficiency of observing. The effective finesse is approximately given
by $1/N^2 = 1/N_r^2 + 1/N_d^2 + 1/N_a^2$, where $N_r$ is the
reflective finesse, $N_d$ is the defect finesse due to plate defects,
and $N_a$ is the aperture finesse due to the solid angle of the beam
(e.g.\ Atherton et al.\ 1981).

FP tunable filters work at low orders of interference to ensure that
the wavelength varies little over the full FOV.  In other
words, they provide quasi-monochromatic images with a single
exposure.  The monochromatic, or Jacquinot, spot is defined to be the
region over which the change in wavelength does not exceed $\sqrt{2}$
times the etalon bandpass, $\delta\lambda$.  From eqn.\ (\ref{eqn1}),
the relative change in wavelength [$\lambda(0)$ --
$\lambda(\theta)$]/$\lambda(0)$ as a function of off-axis angle
$\theta$ is 1 -- cos $\theta$ $\simeq \theta^2$/2 so
\begin{equation}
\frac{\theta^2_{\rm mono}}{2} \simeq \frac{\sqrt{2}\, \delta\lambda}{\lambda}.
\label{eqn2}
\end{equation}
For a wavelength $\lambda$, the bandpass relates directly to the order
$m$ such that $\delta\lambda = \lambda / R = \lambda / Nm$, where $R$
is the spectral resolving power and $N$ is the effective finesse (e.g.
Bland \& Tully 1989).  Combining this expression with eqn.\
(\ref{eqn2}) we find that the angle subtended by the monochromatic
spot is
\begin{equation}
{\theta^2_{\rm mono}} \simeq 2 \sqrt{2}/ (N m)
\label{eqn3}
\end{equation}
For a given order (or, equivalently, via eqn.\ (\ref{eqn1}), a given
plate spacing $l$), the angular size of the monochromatic spot thus
depends on wavelength only through the finesse (the finesse depends on
wavelength through the wavelength-dependent transmission of the plate
coatings).  Eqn.\ (\ref{eqn3}) shows how the spot covers increasingly
larger areas on the detector as the filter is used at lower orders of
interference.

Another notable advantage of FP tunable filters over other
spectrographs is their ability to provide adjustable resolution. Since
$R = N m$, the spectral resolving power, $R$, changes only slowly between orders
when the order number, $m$, is large. One can only get a useful range
of resolutions at low orders, precisely where FP tunable filters
operate.

\section{Description of  MMTF}

\subsection{Host Instrument}

The etalon of MMTF is installed in the collimated beam of IMACS, the
Inamori Magellan Areal Camera and Spectrograph (Dressler et al.\
2006).  IMACS mounts at the Nasmyth focus of the Magellan-Baade
Telescope.  Fed by the f/11 Gregorian configuration, with an integral
atmospheric dispersion corrector (ADC) and field corrector mounted at
the tertiary mirror, the transmitting, all-spherical collimator
produces a well corrected, unvignetted field of 24\arcmin\ in
diameter, and slightly vignetted field of 30\arcmin\ in diameter.  Two
cameras are used to re-image the 150-mm diameter collimator exit pupil
at 0$\farcs$111 and 0$\farcs$200 pixel$^{-1}$ on the detector. To
avoid any major modifications to the design of IMACS and optimize the
FOV, MMTF strictly utilizes the short focal length camera.

The optical path through IMACS and MMTF is shown in Figure 1. The
etalon of MMTF sits in one of six interchangeable positions of the
disperser server (Figure 2).  A mounting plate was fabricated to mate
the Fabry-Perot etalon to the disperser server and facilitate
installation and removal of MMTF.  The CCD array in the f/2
configuration, Mosaic2, consists of a mosaic of 8 E2V 2048 $\times$
4096 devices with 15 $\mu$m pixels.

The MMTF control software is fully integrated into the IMACS control
software.  This aspect of MMTF is described in Section 4 below.

\subsection{Main System Components}

The MMTF consists of 3 major components: the etalon and its support
assembly, the CS100 controller, and the cables connecting them.  The
entire system, except the support assembly, was manufactured by IC
Optical Systems Ltd.

The etalon, shown in Figure 2, consists of two parallel plates
composed of fused silica, each with a 150-mm clear aperture. The
etalons are separated by a sealed but tunable air gap.
The plates are coated on the interior side (the side facing the other
plate) with a reflective coating consisting of a multi-layer
dielectric. The exterior surfaces are anti-reflection coated. The
plates are smooth on small scale at the $\lambda$/100 level.  The
etalon weight is 19.5 kg, not including the mounting plate.

The reflection curve, $\Re(\lambda) \approx 1 - T(\lambda)$ (neglecting
absorption and scattering), of a fused silica witness plate that is
coated in proximity to the etalon plates coating is shown in Figure 3
(the etalon plates and witness are coated simultaneously in the
coating chamber so the only differences between plates and witnesses
are due to distribution in the chamber. This is stated as being 1\% or
less).  The finesse derived from this reflection curve is the
theoretical maximum finesse and is considerably higher than the
effective full-beam finesse of the etalon measured at the telescope
(see Sections 3.4 and 5.1).

The CS100 control system is a three-channel bridge system which uses
capacitance micrometers and long-range stacked PZT actuators,
incorporated into the MMTF, to monitor and correct errors in mirror
alignment and spacing. Two channels (XY) control alignment and the
third (Z) maintains spacing by referencing the cavity length-sensing
capacitor to a high stability standard reference capacitor, also
located in the MMTF. Because this is a closed-loop system,
non-linearity and hysteresis in the PZT drive are eliminated entirely,
as of course are drifts in cavity alignment and spacing. The CS100
servo system works on a nulling method: any imbalance in the
capacitance bridge causing an error signal is corrected for by a
change in piezo voltage and hence plate spacing until the error is
nulled.

The coarse values for tilt and spacing ($X_{\rm coarse}$, $Y_{\rm
  coarse}$, $Z_{\rm coarse}$) are set using dials on the CS100
controller. The tilt and spacing can be changed in smaller increments
using $X_{\rm fine}$, $Y_{\rm fine}$, and $Z_{\rm fine}$ values. The
fine values can be changed using either the CS100 dials or the IMACS
control software (described in Section 4 below). Note that $(X,
Y)_{\rm coarse}$ and $(X, Y)_{\rm fine}$ control the etalon
alignment. However, the coarse spacing $Z_{\rm coarse}$ controls the
etalon bandpass, while the fine spacing $Z_{\rm fine}$ governs the
transmitted wavelength. The fine spacing can affect the bandpass, but
only when changed by a large amount.

The etalon connects to the CS100 through shielded cabling that runs
from IMACS to the cooled electronic rack on one side of the Nasmyth
platform. Since IMACS rotates to compensate for field rotation of the
alt-az mount, the cabling has to be made long enough to accommodate
this rotation. A 3-meter cable connects the etalon to the IMACS
bulkhead via an internal cable wrap. A 15-meter cable connects the
bulkhead to the electronic rack via an external cable wrap (see
Dressler et al.\ 2006 for more detail on the cable wraps in IMACS).

\subsection{Order-Blocking Filters}

A given plate spacing transmits multiple wavelengths at a single
location, each corresponding to a different order (see eqn.\
(\ref{eqn1})), so a blocking filter is required to select only one
transmission order.  The order-blocking filters for MMTF are located
in the f/2 converging beam of IMACS between the camera and the CCD
array (see Figure 1).  The central wavelengths and approximate
full-widths at half-maximum of the transmission profiles (corrected
for the convergence of the IMACS f/2 beam) of the current set of
filters are: 5102/150, 5290/156, 6399/206, 6600/260, 6815/216,
7045/228, 8149/133, and 9163/318 \AA. These filters were selected to
provide access to several emission-line diagnostics (e.g. slightly
redshifted [O~III] $\lambda$5007, [O~I] $\lambda\lambda$6300, 6364,
H$\alpha$, [N~II]$ \lambda\lambda$6548, 6583, [S~II] $\lambda$6716,
6731, and [S~III] $\lambda$9069) and two key long-wavelength
atmospheric OH-free windows centered around $\sim$8150 and $\sim$9150
\AA.

\subsection{Characteristics of MMTF Etalon}

As mentioned in Section 3.2, accurate determination of the
characteristics of the MMTF etalon requires full-aperture illumination
of the etalon in the same configuration (collimated beam) as that used
at the telescope. Table 1 lists a representative summary of on-site
measurements obtained at four different wavelengths (through four
different order-blocking filters). A closer look at the entries in
this table reveals a number of interesting features.

The available full-width at half maximum ({\em FWHM}) instrumental
resolution at 6600 \AA\ is 6 -- 13 \AA.  (As discussed in Jones et
al.\ 2002, the ``effective bandpass'' of FP tunable filters, or the
integral of the profile divided by its peak, is $\pi$/2 $\times$ {\em
  FWHM} rather than 1.06 $\times$ {\em FWHM} in case of a Gaussian
because the instrumental profile is very close to a Lorentzian; see
also Section 4.3.5.) The values of {\em FWHM} at other wavelengths,
for a fixed plate spacing, scale roughly with $\lambda^2$ over 6600 --
9200 \AA, as expected from the expresssion {\em FWHM} = $FSR/N \simeq
\lambda^2 / [2 N l]$, where the effective finesse $N$ is mostly
constant, except at the edges of the etalon's coating response. At the
edges of the coating transmission curve (see Figure 3), the
reflectivity (and hence the etalon finesse) drops, broadening the
profile and increasing {\em FWHM}. Thus, at 5100 \AA, {\em FWHM}
ranges from 8 to 14 \AA. These values are slightly higher than at 6600
\AA, while one would expect lower values for a constant finesse.  The
effective finesse is $\sim$24--29 from $\sim$5300 to 9200 \AA, but
only $\sim$13--14 below $\sim$5300 \AA\ (column (6) in Table 1).

Note that the plate spacings listed in Table 1 represent the effective
distances separating the reflective surfaces, taking into account the
finite thickness of the coatings. It is derived empirically from $l =
\lambda^2/[2 FSR]$. For a fixed physical gap (no change in $Z_{\rm
  coarse}$ and $Z_{\rm fine}$), we find that the effective spacing
depends on wavelength, increasing beyond 6500 \AA\ (this wavelength
dependence is also seen in the effective alignment of the plates; see
Sections 4.3.1 and 5.1).

\subsection{Monochromatic Spot Size}

As described in Section 2 (eqn.\ (\ref{eqn3})), for a given order, the
area of approximately constant wavelength at the image center, or
monochromatic spot, depends on wavelength only through the
finesse. However, the diameter of the spot does depend on plate
spacing via the order number.  The maximum angle of the IMACS
collimated beam is $\sim$10\arcdeg\ at the edge of the short camera
field ($F$ $\simeq$ 365 mm at $\sim$6600 \AA\ with a tendency to
increase at longer wavelengths; see Section 5.1 for more
detail). This is to be compared with the size of the monochromatic
spot derived from eqn.\ (\ref{eqn3}). Table 2 lists representative
values for several different wavelengths and spacings.  Notice the
larger monochromatic spot at 5100 \AA\ due to the lower finesse at
these wavelengths (Figure 3 and Table 1).

\subsection{Sensitivities and Throughput}

The overall sensitivities and throughput of MMTF were calculated at
various wavelengths using continuum spectrophotometric flux standards
(stars from Oke 1990 and Hamuy et al.\ 1992, 1994) and emission-line
flux standards (planetary nebulae from Dopita \& Hua 1997). The
procedure used to flux calibrate MMTF data is described in Section
4.3.5. The results are listed in Table 3. In columns (2) -- (4), we
list the the flux of a 5-$\sigma$ emission-line point source in a
1-hour exposure at New Moon, First or Third Quarter, and Full Moon,
assuming 0.5\arcsec\ seeing, a 1\arcsec\ diameter extraction aperture,
and an airmass of 1.1.  In column (5), we list the throughput in terms
of the ratio of the CCD count rate to the flux.  These measurements
are for optimal plate alignment and assume that the source is a point
source with an emission line centered in the bandpass. For a continuum
source, the flux density in erg s$^{-1}$ cm$^{-2}$ \AA$^{-1}$ can be
inferred by dividing the flux in erg s$^{-1}$ cm$^{-2}$ by the
effective etalon bandpass i.e.\ $\pi$/2 $\times$ {\em FWHM} in \AA\
(see Section 4.3.5 for more detail; the values of {\em FWHM} are given
in Table 1).

\section{Using MMTF}

MMTF has been used for a wide range of applications, including (but
not limited to) the following: imaging of Galactic line-emitting
nebulae and stars, emission-line ratio (``excitation'') maps of
galaxies, searches for warm, diffuse gas on the outskirts of galaxies,
suppression of quasar light in damped Ly$\alpha$ systems, wide-field
surveys for high-redshift line-emitting galaxies. In this section, we
discuss the technical aspects involved in the use of MMTF.

\subsection{Modes of Operation}

MMTF operates in one of three modes: (1) staring mode, (2) wavelength
scanning mode, and (3) charge shuffling and frequency switching
(CS/FS) mode.

In the staring mode, the etalon gap is fixed and the result is a
narrow-band image over the entire FOV. However, the wavelength varies
slowly with distance from the optical axis (eqn.\ (\ref{eqn1})).

In the scanning mode, an image is taken at a series of different
etalon spacings (wavelength slices). The result is a low-resolution
spectrum at each position.  This mode is particularly useful to
capture all line emission from a target which covers a range in
velocity or is larger than the monochromatic spot.

The CS/FS mode enables the observer to switch among 2-3 central
wavelengths during a single exposure. The procedure is as follows (see
Figure 4): Each IMACS CCD is first divided into three equal segments
by an aperture mask. The aperture allows light only onto the center
third of each chip (``open'' segments in Figure 4). When the exposure
begins, the center of each chip is exposed at a single
wavelength. With the shutter closed, the accumulated charge is then
(almost instantaneously) moved down one-third of the chip length. The
etalon's central wavelength is changed (again, almost
instantaneously). The exposure proceeds at this second wavelength,
again at the center of the chip, for a time equal to twice that at the
first wavelength. The charge is then shuffled back up, and integration
proceeds at a third wavelength atop the accumulated charge from the
first part of the exposure for a time equal to that at the first
wavelength. The procedure is then repeated. At the end of the
exposure, each chip contains two images of the same field, each on
one-third of the chip. The last third is empty. The third wavelength
may equal the first, to produce two monochromatic images.
Alternatively, it may be unique from the other two. In this case one
of the resulting images is the sum of images at two separate
wavelengths. Each wavelength in this image experiences one-half the
exposure time of the other image.

The CS/FS procedure allows significant improvements in relative
photometry. Time variations in transparency and seeing are matched
between images at two different wavelengths, using a cadence chosen by
the observer. A second advantage of charge-shuffling is that it allows
accurate matching in wavelength space of the continuum of an
emission-line source. If one image traces an emission-line, the other
can trace the continuum at one or two immediately adjacent
wavelengths.  The time overhead for charge shuffling and changing
wavelengths is limited by the shuffle time: $\sim$2.5 msec per line,
i.e.\ $\sim$ 3 sec for 1300 lines. In comparison the shutter takes
less than 1 second to open or close and the etalon response time is
negligible.  The shutter is closed before the charge is moved on the
chip, and re-opened after the shuffle. The exposure time between
charge shuffles is left up to the observer.  Only 1/3 of the FOV is
illuminated during a given exposure. Thus, for observations of sources
or fields covering the FOV, the camera must be dithered at least twice
on 5\arcmin\ scales to fill in the gaps.  The voltage applied to the
CCD chips for charge shuffling is different than for normal
operation. This results in saturation effects when bright sources are
in the field. Fields containing very bright stars should therefore
be avoided.

\subsection{Installation, Startup, Takedown}

IMACS, the host instrument for MMTF, is by far the most popular
instrument on Magellan-Baade. The MMTF etalon thus cannot stay in the
IMACS disperser wheel when not in use. Although MMTF nights are
generally scheduled in blocks to reduce the number of instrument
switches per semester, an important operational requirement of MMTF
was that installation, startup, and takedown be as straightforward as
possible. At present, the installation, recabling, and startup
procedure takes about an hour from start to finish. The etalon is left
in its mounting plate between the runs to reduce overheads.  The CS100
settings needed to balance the capacitance bridges of the
etalon$\footnote{ The bridges can be unbalanced in several ways. For
  instance, temperature gradients across the MMTF cause changes in
  relative PZT lengths which cause error signals which are nulled by
  changes in the high voltage across the relevant PZT.}$ and align the
etalon plates have proven very stable from run to run. These settings
are used as starting points for the final fine tuning of the alignment
from the control room.

Remote computer control of the CS100 fine settings is done via the
IMACS control software (coarse settings can only be changed
manually). Release and re-acquisition of remote control on the CS100
is done via a button within a password-protected ``HardHat''
engineering GUI. Control of the CS100 is returned to the instrument
front panel if, e.g., the etalon goes out of balance, $Z_{\rm coarse}$
(the bandpass) needs to be changed, or the CS100 is being turned off
at the end of a run.

MMTF observing is done using the standard IMACS data acquisition
software along with a set of special built-in procedures. These
include the ability to set the etalon fine X, Y, and Z values in the
Hardhat GUI, as well as a set of programmable scripts which coordinate
commands to the CCD camera and the CS100 for the following calibration
and observation tasks: (1) wavelength scanning; (2) CS/FS; (3)
determining the plate alignment; (4) taking ``data sausages'' for
wavelength calibration. We explain these tasks in the following two
sections, starting with the calibration tasks followed by the
observation tasks.

\subsection{Calibrations}

Full calibration of MMTF consists of 5 different tasks: (1) aligning
the etalon plates, (2) checking the position of the optical axis, (3)
determining the wavelength calibration, (4) flat fielding the data,
and (5) deriving absolute photometry. The first four of these can be
carried out by the instrument scientist and/or the observer before the
night begins. These tasks are carried out within the IMACS control
software by creating standard scripts and running them.

\subsubsection{Plate Alignment}

Alignment is the process by which the two etalon plates are made
parallel to one another. If the plates are not optimally aligned, the
transmission profile will be broad and/or asymmetric. This diminishes
the light transmitted in the core of the profile, and thus the system
throughput (for an emission-line source) and the instrumental
resolution. It is thus crucial that proper alignment is achieved and
maintained throughout the run.

There are several possible methods for achieving parallel plates,
providing various degree of accuracy (e.g.\ Jones \& Bland-Hawthorn
1998).  We have developed a simple and efficient procedure which
relies on scanning the transmitted image in the radial coordinate to
produce a spectrum. The plate alignment is scanned along both axes of
movement using a (M x N) grid of [$X_{\rm fine}, Y_{\rm fine}$]. At
each value of [$X_{\rm fine}, Y_{\rm fine}$], an 1-5 sec exposure is
taken of a bright emission line from an arc lamp appropriate for the
order-blocking filter being used (Argon, Krypton, Neon, or Xenon).
Each image is then azimuthally averaged to create an emission-line
spectrum (Figure 5; see Section 4.5.3 for more detail on the method
used to create this spectrum). The profiles of emission features are
compared by eye (and, where possible, with line fits) to find the
narrowest and most symmetric profile. An example of output from this
procedure is shown in Figure 6. This alignment procedure typically
takes $\sim$15 minutes from beginning to end.

As briefly described in Section 3.2, plate alignment is achieved and
maintained through a capacitor and piezo-electric feedback
system. However, we discovered that plate alignment unexpectedly
depends on two variables: (1) the wavelength, due to changes in the
behavior of the multi-layer reflection coating with wavelength; and
(2) the rotation angle of the etalon, due to gravity-induced sag of
the plates that remains uncorrected by the system feedback mechanism.
These dependencies are shown in Figures 7 and 8, respectively.
Alignment changes by $\la$ 0.02 $\mu$m from 5100 to 6500 \AA\ but by
$\sim$0.10--0.15 $\mu$m from 6500 to 9100 \AA. The total amplitude of
the change in alignment with gravity angle is $\sim$0.03 $\mu$m, but
is generally highly reproducible.$\footnote{For reasons which remain
  unclear, the alignment values become difficult to predict for
  gravity angles between $-$120 and $-$20 degrees, a region dubbed the
  ``danger zone''.}$

The plate alignment therefore has to be adjusted whenever the
order-blocking filter is changed. The change in alignment due to
rotation is minimized by breaking up long exposures into segments of
$\sim$20-30 minutes and starting each exposure at approximately the
same IMACS rotation angle (gravity angle). During each exposure, the
IMACS camera rotates to maintain a constant orientation of the CCD
mosaic with respect to the sky, but the change in gravity angle over
the course of these short exposures is small enough to have no
appreciable effect on plate alignment (this issue is discussed further
in Sections 4.4 and 5.1).

\subsubsection{Optical Axis}

There is a wavelength gradient in the MMTF due to the angle at which
rays pass through the etalon (eqn.\ (\ref{eqn1})). This gradient is
circularly symmetric about the optical axis (the projection of the
MMTF normal axis onto the CCD). The optical axis is in the center of
the FOV, but shifts slightly from run to run (due to the
re-mounting of the IMACS CCD) and must be re-measured. During a
particular run, it is constant within a few pixels as long as the
etalon plates stay aligned. 

To find the optical axis, we use the fact that there is a faint
($\sim$0.5\%) ghost reflection between the MMTF and the CCD (this is a
well-understood feature of FPs; Bland \& Tully 1989). Once the etalon
plates are properly parallel, we insert into the beam a slit mask that
has 5-10 guide star holes (square apertures) dispersed somewhat evenly
over the central half of the FOV, and then take a quartz lamp
exposure.  The exposure must be long enough that faint reflections
(primary ghost images) of the guide star boxes appear, but short
enough that the original images do not saturate. These reflections are
symmetric about the optical axis. An IRAF procedure is run to
determine the positions of the guide star boxes and their reflections
and then derive the location of the optical axis, typically with
accuracy of $\la$1 pixel ($\la$0$\farcs$2). An example of output from
this procedure is shown in Figure 9. This procedure typically takes
$\sim$15 minutes from beginning to end.

\subsubsection{Wavelength Calibration}

On the optical axis, for effective plate spacing $l$, $\lambda = A + B
Z_{\rm fine}$, where $Z_{\rm fine} = Z(l)$ represents voltages applied
to the etalon by the CS100 and is the variable read by the IMACS
computer. It is a linear function of plate spacing $l$: $l = E + G
Z_{\rm fine}$, where $E = m A / 2$ and $G = m B / 2$. $A$ and $B$ thus
vary from order to order. By substituting for $m$ using the
interference equation, eqn.\ (\ref{eqn1}), we also see that $A$ and
$B$ vary linearly with wavelength for a given plate spacing. More
generally, using eqn.\ (\ref{eqn1}) at an arbitrary radius $R$ off the
optical axis:
\begin{equation}
\lambda(R) = \lambda(0) / \sqrt{(1 + R^2/F^2)} = (A + B Z_{\rm fine})
/ \sqrt{(1 + R^2/F^2)},
\label{eqn4}
\end{equation}
where $F$ is the focal length of the camera (note that the wavelength
decreases with increasing distance from the optical axis.).

Wavelength calibration involves finding $A$ (the wavelength scale zero
point) and $B$ ($\equiv d\lambda/dZ$) in eqn.\ (\ref{eqn4}).  $A$ may
vary with time (possibly due to temperature changes) and with the
alignment setting. It is calibrated at the beginning of the run and
checked frequently using single, full-field images of emission-line
rings from an arc lamp (Figures 5 and 10; also see Section 4.4 for a
discussion of $A$ drift monitoring). $B$ depends on the coarse plate
spacing, $Z_{\rm coarse}$, and the interference order. It is stable
and has been measured for 3 values of $Z_{\rm coarse}$ in each filter
based on MMTF arc lamp spectra (Figure 11). These were extracted from
``data sausages'' ({\em i.e.} data cubes which are narrow in the X and
Y directions but highly elongated in the Z direction). These sausages
were obtained by scanning the etalon spacing through somewhat more
than one free spectral range and using small CCD subrasters at 8
different distances from the optical axis (one on each CCD chip),
including one near the center where the main science target is often
positioned. It currently takes about 20-30 minutes to acquire these
Nyquist-sampled data sausages.  These spectra are compared with fully
calibrated archived spectra obtained with the same settings to help
with line and order identifications.  The fit takes into account all
eight data sausages and provide the $A$ and $B$ parameters for all
orders that were sampled.

\subsubsection{Flat Fielding}

Flat fielding corrects for large-scale and pixel-to-pixel variations
in sensitivity across the field. The amplitudes of these variations
depend on wavelength so a series of flats should in principle be
taken at each wavelength in which the science targets were observed.
In practice, we have found that steps of $\sim$10 \AA\ are sufficient
to correct for this wavelength dependence. A screen in front of the
secondary mirror is illuminated by a bright quartz lamp and exposure
times of 3-5 seconds are sufficient to get $\sim$1\% accuracy.  In the
case of observations taken in the CS/FS mode (Section 4.1), the flats
are taken in the exact same mode, adjusting the exposure times
appropriately to prevent saturation.

\subsubsection{Absolute Photometry}

Absolute flux calibration with MMTF is straightforward. A
spectrophotometric standard of moderate brightness (e.g. Oke 1990;
Hamuy et al.\ 1992, 1994) is observed at roughly the same wavelength
at which the science data were taken. The calibration star is
positioned roughly at the same location as the science target, since
wavelength changes with distance from the optical axis.  The
transmitted stellar flux from the standard is $f_\lambda \times$
``effective bandpass'', where $f_\lambda$ is the continuum flux
density of the star at wavelength $\lambda$ and the effective bandpass
is the integral of the profile divided by its peak. As discussed in
Section 3.4, the effective bandpass for the Lorentzian instrumental
profile of MMTF is $\pi$/2 $\times$ {\em FWHM} rather than 1.06
$\times$ {\em FWHM} in case of a Gaussian (see Jones et al. 2002 for
more detail).

Care must be exercised in choosing a stellar flux standard where the
wavelengths of interest are not affected by stellar absorption lines
(e.g.\ H$\alpha$).  To avoid this problem altogether, emission-line
flux standards (e.g.\ planetary nebulae of Dopita \& Hua 1997) are
often used when observing in the 6600 \AA\ and 6815 \AA\
order-blocking filters.  These emission-line stardards are observed in
wavelength scanning mode (see Section 4.1 for a description) to
produce integrated emission-line spectrum that can be directly
compared with the known H$\alpha$, [N~II] $\lambda\lambda$6548, 6583 or
[S~II] $\lambda\lambda$6716, 6731 fluxes of these objects.

\subsection{Observations}

\subsubsection{Before the Run}

Three options are available to allow the observers to customize MMTF
to their science program:

\noindent{\em 1. Choice of observing mode.} As described in Section
4.1, MMTF can be used in three different modes. The staring mode is
ideal for blind searches of emission-line sources over large
areas. The scanning mode is useful to increase the volume of
stare-mode blind searches or to study sources with a range of
velocities or of angular size larger than the monochromatic
spot. Finally, the CF/FS mode is well suited for high-precision
differential experiments which require only a small fraction of the
total FOV. 
 
\noindent{\em 2. Choice of spectral resolution.} The choice of $Z_{\rm
  coarse}$, or coarse plate spacing, is a compromise between the size
of the monochromatic spot and the signal-to-sky contrast. Larger
values of $Z_{\rm coarse}$ (higher orders) yield smaller monochromatic
spot sizes (eqn.\ (\ref{eqn3})). However, they also have smaller
bandpasses, which means that the amount of sky emission transmitted is
lower, thus increasing the signal-to-noise ratio.  Three $Z_{\rm
  coarse}$ settings are available to the observer; it is generally not
changed during an observer's run due to overheads associated with
the calibrations.

\noindent{\em 3. Choice of wavelength and order-blocking filters.} 
The observer is limited to the use of at most 2 filters in a single
night because of calibration overheads.  Observers have full freedom
to tune the central wavelength of the MMTF bandpass within the range
allowed by a particular blocking filter. The wavelength can be tuned
from one exposure to the next, or within a given exposure (see Section
4.1).  Changing filters between exposures is straightforward. However,
each filter requires its own set of calibrations to determine the
wavelength solution and etalon alignment (Section 4.3).

The MMTF website has a number of GUI tools to help with proposal
writing, run planning, and the observing itself: (1) a transmission
calculator is available to determine if the wavelength of a particular
redshifted feature falls within the order-blocking filters of MMTF and
the corresponding transmission of the relevant filter.  (2) a exposure
time calculator has recently been implemented to determine the
exposure time needed to reach a particular signal-to-noise ratio or,
conversely, the signal-to-noise ratio reached for a particular
exposure time. The wavelengths of strong OH lines that fall within the
observed range are flagged to warn the users that the signal-to-noise
ratio at these wavelengths will be degraded due to enhanced Poisson
noise and larger systematic errors resulting from sky subtraction. (3)
A wavelength calculator has been designed to help determine the exact
position of emission lines on the detectors for a particular MMTF
etalon setting based on a previously determined wavelength
calibration. This last calculator is also used during the night to
update the $A$ value of the wavelength calibration in case of drifts.

\subsubsection{During the Afternoon}

For each filter to be used during the night, the following steps are
taken in the afternoon preceding the night: (1) aligning the etalon
plates using the profiles of an emission line from an arc lamp
(Section 4.3.1), (2) capture a binned reference image of an
emission-line ring from an arc lamp (Section 4.3.3), and (3) take
the data sausages of an arc lamp over more than one FSR (Section
4.3.3). These calibrations are generally performed at a gravity angle
of $\theta_{\rm grav} \sim 0^\circ$, the parked position of IMACS
during the day. The results from the wavelength calibration ($A, B$)
are inserted in the wavelength calculator for use throughout the
night.

\subsubsection{At the Start of the Night}

The first target is acquired at the same IMACS gravity angle as the
calibrations to avoid shifts in plate alignment due to etalon
rotation. Before the observation, an exposure is taken of an arc lamp
using the same $Z_{\rm fine}$ setting and on-chip binning as those
used for the reference image. This image is azimuthally averaged to
create an emission-line spectrum which is then compared with the
afternoon reference image to search for possible shifts in the zero
point of the wavelength calibration.  The value of $A$ in the
wavelength calculator is updated accordingly. Using these updated
numbers, the appropriate $Z_{\rm fine}$ for the wavelength of interest
is computed and the etalon is set at this value.

\subsubsection{Throughout the Night}

Before each exposure, IMACS is rotated to match the gravity angle of
the calibrations.  This procedure is repeated before each exposure of
a sequence to minimize errors in the wavelength solution. We have
found that this ``fixed gravity angle procedure'' cuts down on nightly
overheads by up to 1--2 hours.

A reference ring is taken regularly to check for change in peak
position (wavelength zero-point, $A$) and line profile (plate
alignment). This is done every $\sim$20--30 minutes while the
temperature is changing quickly at the start of the night, or every
hour when the temperature is stable. If the line profiles have
degraded (less narrow and/or less symmetric), the alignment is
recomputed using the procedure described in Section 4.3.1 and a
reference ring is re-acquired to update the value of $A$ and
recalculate the value of $Z_{\rm fine}$ which corresponds to the
wavelength of interest (Figures 6 and 10).

We have found that realignment of the plates is generally not needed
during the night if the fixed gravity angle observing procedure is
used.  This procedure does not allow for exposure times longer than
20-30 minutes.  Nevertheless, each of these relatively short exposures
is sky-noise limited so there is no loss of signal-to-noise ratio when
the multiple exposures of a sequence are combined together.

Dithering of 30$\arcsec$ or more in each direction is recommended to
achieve a continuous image uninterrupted by features or chip gaps.
Dithering is important for IMACS imaging in order to bridge the
inter-chip gaps, which are $\sim$50 pixels, or 10$\arcsec$, on
average.  MMTF also introduces a set of reflections of the chip edges
(of relative intensity of a few percent) which are not corrected by
the flat fields.  These reflections contaminate $\sim$10 rows and/or
columns of each chip. 

\subsubsection{At the End of the Night}

Flats are obtained for each setting separated by at least $\sim$10
\AA\ observed during the night (Section 4.3.4).  These flat field
exposures have been found to be quite stable with time so they are
acquired either in the morning after the observing or the afternoon
prior the observing. A series of biases (0-second exposures) is
obtained after the flats.

\subsection{MMTF Data Reduction Pipeline}

We have developed a collection of Fortran routines and Perl-driven
IRAF scripts to calibrate the MMTF, subtract sky background from data,
and create annular image masks. This software is permanently installed
on the computers at Magellan, but is also publicly available on the
MMTF website for the user's convenience when doing calibrations and
data reduction elsewhere. 

The main steps in the reduction pipeline are: (1) bias subtraction and
flat fielding, (2) cosmic ray and bad pixel masking, (3) sky
subtraction, (4) astrometry registration and image mosaicking, and (5)
PSF-matching and stacking.  The data from IMACS emerge as 8 individual
files for each exposure, one file per chip. Steps 1-4 operate on these
individual files, while Step 5 operates on the mosaic. It takes
approximately one day to reduce a night of MMTF data on a 2.8 GHz
machine with 2 Gb of memory. The most time-consuming step is
the cosmic ray removal. Each step is briefly described below.

\subsubsection{Bias Subtraction and Flat Fielding}

Bias subtraction and flat fielding of MMTF data are straightforward:
The row overscan is subtracted from all files.  The biases are
combined.  The combined bias is subtracted from each source and flat
field file.  The column overscan is subtracted.  The flats are
combined.  Each flat is normalized so that the central region of the
FOV has an average value near unity. The same normalization factor is
applied to each chip. If specified, scattered light signatures are
removed from the flat field. The source files are divided by the
normalized flat field exposure.

\subsubsection{Cosmic Ray and Bad Pixel Masking}

Cosmic ray removal is achieved in two stages. First, we attempt to
remove cosmic rays for each image individually. This is done using
Pieter van Dokkum's {\em ``LA Cosmic''} task for IRAF (for details,
see http://www.astro.yale.edu/dokkum/lacosmic/). This task is called
from a script, which has switches to allow control of such things as
the sky level and the threshold sigma for rejection. Generally, four
iterations is enough to remove the majority of the cosmic rays in each
field.  Known bad pixel regions (e.g.\ bad columns) are appended to
the cosmic ray masks. Although a very effective task, {\em LA Cosmic}
tends to leave residual bright pixels unmasked (although very rarely
does it miss a cosmic ray completely). This is remedied by using the
IRAF {\em ``crgrow''} task later on to expand the masks by a single
pixel in each direction (Section 4.5.4) and filter out any residual
cosmic rays using the {\em ``imcombine''} during the stacking
procedure (Section 4.5.5).

\subsubsection{Sky Subtraction}

Sky subtraction is performed using the MMTF calibration
software. Given the knowledge of the mapping of the CCD pixels to the
focal plane and the location of the optical axis, the sky spectrum is
azimuthally averaged. Sources and cosmic rays are filtered out using a
biweight statistic. If necessary, median smoothing is employed to
smooth out the sky spectrum. This results in a robust removal of the
sky using a single command. This procedure runs without user
interaction. On a 2.8 GHz machine with 2 Gb of memory, the sky
subtraction takes approximately 5-10 minutes per exposure (for
unbinned data).  If the optical axis is not accurate to within
$\sim$20 pixels or so, dipole residuals appear after sky subtraction,
either in individual exposures or in a stack of several exposures with
small dithers. If the etalon plates are properly parallel, the optical
axis position derived from the method described in Section 4.3.2 will
be correct well below this level. For exposures with large, extended
sources in the FOV, a subset of the CCDs is selected to avoid bright
source emission in the calculation of the sky spectrum.

\subsubsection{Astrometry Registration and Image Mosaicking}

Perhaps the most challenging aspect of IMACS data reduction is
determining the astrometric solution that maps the physical CCD
coordinates to locations on the sky. Because of the small focal ratio
of the IMACS camera (f/2) when the MMTF is in use, as well as the wide
FOV, significant deviations from a linear mapping occur.  The
higher-order polynomial terms of the FOV-to-sky mapping are known from
observations of dense star fields. The script applies the World
Coordinate System (WCS) information to all 8 chips for each
exposure. This information contains the high-order distortions in the
chips as well as the accurate chip spacings and the rotation and scale
of the FOV.  This information is used to properly register each FOV.

Cosmic ray masks are extended by one pixel in each direction to
prevent ringing caused by residual cosmic rays in the interpolation
steps. Masked pixels in the input files are replaced by 0.0 (the sky
value) to prevent blurring of cosmic rays. These values will not play
a role in the final image, since the contribution from masked images
will be ignored. The only reason for this coarse interpolation is to
prevent ringing in the next step.

The IRAF task {\em ``mscred.mscimage''} is run on each individual chip
in order to apply the high-order tangent plane projection (TNX)
corrections. The bad pixel mask undergoes the same transformation as
the image. This task is run on each chip individually because {\em
  ``mscimage''} will not apply TNX corrections on mosaicked images
(this is a known IRAF issue).

The chips are combined into a composite Multi-extension FITS (MEF)
file and {\em ``mscimage''} is run a second time to create a single
file with a single bad pixel mask including cosmic ray, bad column and
chip gap locations.  A catalogue of stellar positions is created using
the IRAF task {\em ``mscred.mscgetcat''}. Stars falling anywhere
within roughly 3500 pixels from the center and with magnitudes ranging
from 19 to 10 are used.  The IRAF task {\em ``mscred.msccmatch''} is
used to match the stellar positions to pixel positions in order to
determine the lower-order corrections for the mosaicked image. These
corrections are different from exposure to exposure due to various
atmospheric effects and pointing errors. These small corrections are
also applied to the bad pixel masks. Using the catalogue of stellar
positions, a catalogue of PSFs are made for each FOV. This is used in
the combining stage when PSF matching is required (next section).

\subsubsection{PSF-Matching and Stacking}

Before combining dithered exposures into a single, deep image, the
various image groups are PSF-matched and aligned as follows.  Using
the PSF catalogues produced by {\em ``imacsreg''}, the mode seeing
value is computed for each exposure and inserted into the header.  The
images are aligned based on the WCS information. These offsets account
for any dithering or pointing errors. Bad pixel masks are also shifted
accordingly. All similar frames are degraded to the worst seeing using
the IRAF task {\em ``psfmatch''} (making sure to remove frames with
anomalously bad seeing). Similar frames are stacked using the {\em
  ``imcombine''} task in IRAF. The process of combining images ignores
masked pixels and also rejects all values above a certain threshold
(e.g.\ 2 sigmas), thus removing any residual cosmic rays.

\section{Lessons learned}

The commissioning and implementation of MMTF as a facility instrument
have been harder than anticipated.  We highlight here the main lessons
learned along the way.


\subsection{Technical Issues}

\noindent{\em 1. The etalon alignment depends on position angle.}
Detailed testing of MMTF revealed that the alignment of the MMTF
etalon depended sensitively ($\pm$ $\sim$0.03 $\mu$m; Figure 8) on the
rotation angle of the etalon with respect to the (horizontal) optical
axis of the Nasmyth focus. This effect is not seen when the optical
axis is near vertical, e.g.\ at Cassegrain focus. This effect is
largely reproducible with position angle (Figure 8) and is attributed
to the sagging of the large heavy plates of the MMTF etalon under
gravity. It is now dealt with by forcing IMACS (and thus MMTF) to
return to roughly the same gravity angle for each observation. This
observing strategy has eliminated the need for time-consuming
re-alignment and reduced the drift in the wavelength solution.
However, it causes the field to rotate with respect to the CCD axes
from observation to observation, so data reduction has to de-rotate
the field before stacking a sequence of observations of the same
target. We have found that this procedure does not affect the image
quality of the final stack to within a fraction of a pixel even at the
outer edge of the field.

\noindent{\em 2. The etalon alignment depends on $\lambda$.}
We also discovered that the $(X, Y)$ settings needed to align the
etalon plates for a given etalon rotation angle were significantly
different at longer wavelength (Figure 7). This means that the
effective reflecting surfaces at long wavelengths are wedged with
respect to the surfaces at shorter wavelengths, and vice-versa.  We
were expecting the effective gap (measured from $l = \lambda^2/[2
FSR]$) to vary with wavelength for a fixed gap (and this has been
shown to be the case; Table 1; see also Rangwala et al.\ 2008), but we
were not expecting the {\em relative angle} between the effective
reflecting surfaces to vary this much with wavelength. This effect was
dealt with by simply re-aligning the etalon plates after each change
of order-blocking filter.

\noindent{\em 3. The etalon alignment does not depend on temperature.}
The CS100 $X$, $Y$ settings needed to align the etalon plates were
found to be very stable from run to run over a period of three years,
given the same cables, CS100, etalon rotation angle, and wavelength
were used. This implies that the etalon alignment is not sensitive
to temperature changes. This reduces overheads associated with the
installation and startup of MMTF.

\noindent{\em 4. The position of optical axis depends on etalon
  alignment.}
We have found that it is important to align the etalon plates before
deriving the position of the optical axis. 

\noindent{\em 5. The connectors to the cables are ``fickle''.}
MMTF uses a set of 15, 3, and 1 m cables provided by IC Optical
Systems to connect the etalon to the CS100 controller. These long
cables were covered by a copper sheath to shield them from the
telescope electronic noise. These cables need to be handled with
care. This is particularly true of the connectors which have a
tendency to loosen easily. Applying a small drop of screw locking
compound or cyanoacrylate glue (super glue) is often a good cure for
this problem.

\noindent{\em 6. The focal length of the f/2 camera depends on
  $\lambda$ and $Z_{\rm coarse}$.}  The empirically derived focal
  length of the f/2 camera was found to depend slightly on wavelength
  and coarse etalon spacing ($Z_{\rm coarse}$) (see Figure 12).  The
  wavelength dependence indicates slight chromatic aberrations due to
  IMACS, MMTF, or both. The change in focal length is steeper beyong
  $\sim$6600 \AA, roughly coincident with the change in etalon
  alignment (Section 5.1.2, Figure 7). We therefore speculate that
  MMTF (more specifically its coatings) is at least partly responsible
  for the chromatic aberrations. The dependence of the focal length on
  etalon spacing (at fixed wavelength) adds support to this idea. We
  calibrated these dependences and modified the wavelength calculator
  to take this effect into account.

\noindent{\em 7. The CS/FS mode of MMTF is of specialized use.}
The ``active area'' of the detector in the CS/FS mode is at 7\arcmin\
or more from the optical axis so the monochromatic area in this mode
is a ring with only {\em FWHM} $\approx$2$\farcm$1 or less. This mode
is therefore only good for relatively small objects and does not take
advantage of the large FOV of MMTF.

\noindent{\em 8. The full-field finesse of the MMTF etalon is lower
  than expected.}
Irregularities in the plate surfaces and coating thicknesses, as well
as deviations from perfectly parallel plates, degrade the effective
finesse (hence the efficiency of observing), from a value of $\sim$50
expected from the reflectivity curve (Figure 3) to a value of
$\sim$24--29 (Table 1).

\subsection{Scientific Issues}

\subsubsection{Strengths of MMTF}

Our experience has helped us appreciate the scientific strengths and
weaknesses of MMTF. One outstanding feature of MMTF is its very large
FOV (27\arcmin\ diameter), making it an ideal survey
instrument, particularly if monochromaticity is not a critical
factor. And, even if it is, the large Jacquinot spot (up to
$\sim$13\arcmin\ in low resolution mode; Table 2) of MMTF makes it a
very powerful instrument which achieves a significant efficiency
advantage over conventional long-slit spectrographs (Jacquinot 1954).

The large aperture of the Magellan-Baade telescope, superb delivered
image quality (DIQ $\sim$0$\farcs$5) of the IMACS f/2 camera,
excellent photometric characteristics of the Las Campanas site, and
ability of MMTF to tune the transmitted wavelength to any feature
within the range $\sim$5000--9200 \AA\ combine to provide excellent
emission-line sensitivity to point sources or filamentary structures
(see Table 3) and make MMTF a highly versatile instrument.

Through rejection of significant sky and/or continuum emission
(compared to conventional narrow-band filters), MMTF can significantly
improve observing efficiency for reaching a desired signal-to-noise
ratio. The gain is particularly high in spectral regions where OH sky
lines are densely distributed ($\ga$ 7000 \AA).

MMTF can switch between two wavelengths on short timescales (e.g.
on-band and off-band images obtained in sequence or nearly
simultaneously through the use of the CS/FS mode). This allows the
user to correct for time-varying observational effects, including
atmospheric transmission and sky brightness. It also allows for
differential imaging on short timescales.

FP tunable filters in general are unbeatable for wide-field diffuse
light detection (no IFUs can ever compete in this area of research),
although this type of program does not take advantage of the
excellent DIQ of MMTF/IMACS.

\subsubsection{Weaknesses of MMTF}

Contrary to TTF (e.g.\ Bland-Hawthorn \& Kedziora-Chudczer 2003), MMTF
has proven difficult to use in bright moon conditions.  Scattered
light in the f/2 camera and MMTF optics makes data obtained in these
conditions tricky to reduce.  Observations under grey or darker
conditions do not show these problems. 

Internal scattering also produces complex halos around bright stars.
The shape of the halos and their integrated intensity relative to that
of the star ($\sim$10\%) vary across the field so they are difficult
to model. Targets near very bright stars have been observed by putting
the star within the chip gap.

MMTF is not a good instrument for programs requiring multiple ($>$2)
blocking filter switches in one night.  Each filter used requires a
set of careful wavelength calibration during the day ($\sim$1 hour
each; see Section 4.4.2 for more detail). Changing filters in the
middle of the night adds an overhead of 15-30 minutes due to these
calibrations.

MMTF is better suited for deep observations of a few targets per night
rather than snapshots of a large number of targets due to the overhead
associated with the rotation of IMACS for each target.

\section{Outlook}

MMTF is currently used at a rate of $\sim$10 nights per semester.  As
MMTF becomes a facility instrument and a community of users is
created, we anticipate that this rate will increase. 

To our knowledge, two similar instruments on 10-meter class telescopes
will soon be in routine operation: the imaging Fabry-Perot system for
the Robert Stobie Spectrograph on the South African Large Telescope
(Rangwala et al.\ 2008) and OSIRIS, the Optical System for Imaging and
Low-Intermediate-Resolution Integrated Spectroscopy on the Gran
Telescopio Canarias (Cepa et al.\ 2007 and references therein). These
instruments will cover a different area of the parameter space than
MMTF, taking advantage of the larger collecting area of their host
telescopes but over a smaller FOV than MMTF. As they mature, the
combination of these three tunable filters will provide formidable
tools to study emission-line systems, near and far. MMTF will remain
the widest field device of its kind in the foreseeable future.

\acknowledgements S.V., B.J.W., D.S.N.R., and M.M. were
supported in part by NSF through contracts AST/ATI 0242860 and AST/EXC
0606932.  S.V.\ acknowledges support from a Senior Award from the
Alexander von Humboldt Foundation and thanks the host institution, MPE
Garching, where this paper was written.  The MMTF team is grateful for
the support provided for this project by the past and present
Directors of the Observatories of the Carnegie Institution for
Science, Gus Oemler and Wendy Freedman, the Magellan Technical
Manager, Alan Uomoto, the Magellan Council, and the technical staff at
the Las Campanas Observatory. We also thank P.L.\ Shopbell and A.\
Bagish who helped in the early stages of the MMTF testing and
commissioning, and M.\ Bergamo who helped put together the GUI for the
MMTF Exposure Time Calculator. Last but not least, we are very
grateful to R.J.\ Weymann who, more than eight years ago, helped
initiate this project which is now a reality.

\clearpage

\begin{deluxetable}{cccccc}
\tablecolumns{6}
\tabletypesize{\scriptsize}
\tablecaption{Characteristics of the ET-150 Etalon in MMTF}
\tablewidth{0pt}
\tablehead{
\colhead{$Z_{\rm coarse}$} & \colhead{$m$} & \colhead{$l$ ($\mu$m)} & 
\colhead{$FWHM$ (\AA)} & \colhead{$FSR$ (\AA)} & \colhead{$N$}\\
\colhead{(1)} & \colhead{(2)} & \colhead{(3)} & \colhead{(4)} &  \colhead{(5)} & \colhead{(6)}
}
\startdata
\cutinhead{$\lambda$ = 5102 \AA}
--2 & 27.5 & 7.0 & 13.7 & 186 & 13.6\\
+1 & 37.7 & 9.6 & 10.1 & 135 & 13.4\\
+3 & 47.6 & 12.1 & 8.0 & 107 & 13.4\\
\cutinhead{$\lambda$ = 6600 \AA}
--2 & 20.0 & 6.6 & 11.8 & 314 & 26.6\\
+1 & 29.0 & 9.6 & 8.1 & 219 & 27.0\\
+3 & 35.0 & 11.6 & 6.6 & 181 & 27.4\\
\cutinhead{$\lambda$ = 7045 \AA}
--2 & 22.8 & 8.3 & 10.3 & 299 & 29.0\\
+1 & 30.8 & 11.1 & 7.7 & 224 & 29.1\\
+3 & 35.8 & 13.0 & 6.6 & 191 & 28.9\\
\cutinhead{$\lambda$ = 9163 \AA}
--2 & 22.0 & 10.2 & 17.2 & 414 & 24.1\\
+1 & 29.0 & 13.5 & 13.1 & 314 & 24.0\\
+3 & 33.0 & 15.3 & 11.5 & 276 & 24.0\\
\enddata
\tablecomments{ Col.\ (1): Coarse $Z$ setting on etalon. Col.\ (2):
  Order number.  Col.\ (3): Plate spacing in $\mu$m. Col.\ (4):
  Spectral resolution in \AA.  Col.\ (5): Free spectral range in
  \AA. Col.\ (6): Effective finesse = {\em FSR/FWHM}.}
\label{tab:et150}
\end{deluxetable}

\begin{deluxetable}{ccc}
\tablecolumns{3}
\tabletypesize{\scriptsize}
\tablecaption{Jacquinot Spot Diameters}
\tablewidth{0pt}
\tablehead{
\colhead{$\lambda$(\AA)} & \colhead{$D_J$ (Zcoarse = --2) (arcmin)} & \colhead{$D_J$ (Zcoarse = +3) (arcmin)} \\
\colhead{(1)} & \colhead{(2)}  & \colhead{(3)} 
}
\startdata
5102	& 13.8\tablenotemark{a}	& 10.7\tablenotemark{a}\\
6600	& 11.6	& 8.5\\
9163	& 11.5	& 9.5\\
\enddata
\tablecomments{ Col.\ (1): Central wavelength. Col.\ (2): Jacquinot spot
  diameter at Zcoarse = --2.  Col.\ (3): Jacquinot spot diameter at
  Zcoarse = +3.  }
\tablenotetext{a}{At 5100 \AA\ the numbers are $\sim$1--2\arcmin\ higher
  due to a drop in finesse.}
\label{tab:splot}
\end{deluxetable}

\begin{deluxetable}{ccccc}
\tablecolumns{5}
\tabletypesize{\scriptsize}
\tablecaption{MMTF Detectable Flux and Throughput}
\tablewidth{0pt}
\tablehead{
\colhead{$\lambda$} & \colhead{Detectable Flux} & \colhead{Detectable Flux} & \colhead{Detectable Flux} & \colhead{Throughput}\\
\colhead{(\AA)} & \colhead{(10$^{-17}$ erg s$^{-1}$ cm$^{-2}$)} & \colhead{(10$^{-17}$ erg s$^{-1}$ cm$^{-2}$)} & \colhead{(10$^{-17}$ erg s$^{-1}$ cm$^{-2}$)} & \colhead{(10$^{16}$ counts erg$^{-1}$ cm$^{2}$)} \\
\colhead{(1)} & \colhead{(2)} & \colhead{(3)} & \colhead{(4)} & \colhead{(5)} 
}
\startdata
5102 & 2.90 & 3.27 & 4.44 & 0.11\\
6600 & 0.66 & 0.75 & 1.04 & 1.42\\
6815 & 0.40 & 0.45 & 0.61 & 1.76\\
7045 & 0.52 & 0.59 & 0.78 & 1.42\\
8149 & 1.98 & 2.09 & 2.47 & 0.18\\
\enddata
\tablecomments{ Col.\ (1): Wavelength.  Col.\ (2): Flux of a
  5-$\sigma$ emission-line point source in a 1-hour exposure, assuming
  Moon age = 0 days, 0.5\arcsec\ seeing, 1\arcsec\ diameter extraction
  aperture, and airmass of 1.1. Col.\ (3): Same as Col.\ (2) but Moon
  age = $\pm$7 days. Col.\ (4): Same as Col.\ (2) but Moon age = 14
  days. Col.\ (5): Ratio of count rate to flux.}
\label{tab:sensitivities}
\end{deluxetable}

\clearpage

\setcounter{figure}{0}
\begin{figure*}[ht]
\epsscale{1.0}
\plotone{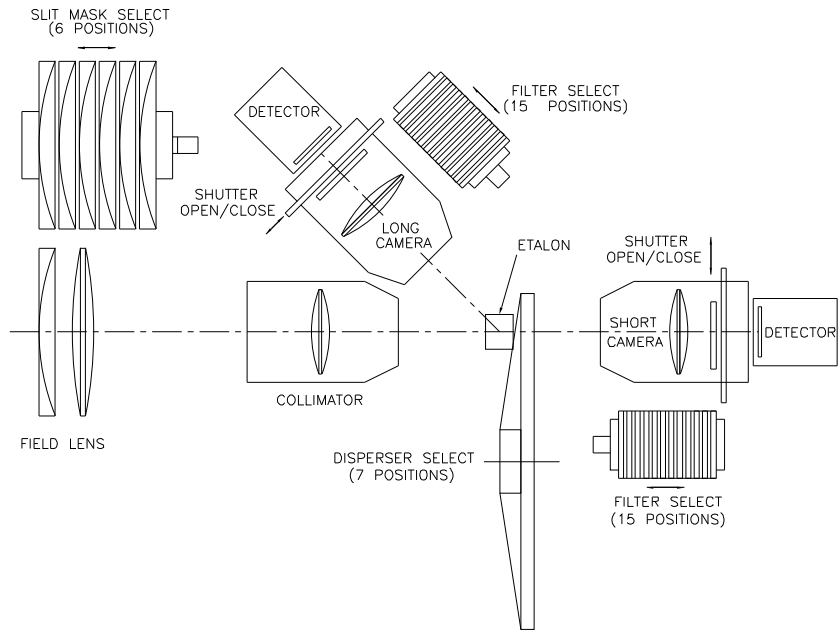}
\caption{ The optical path through IMACS and the MMTF. The MMTF etalon
  is placed in the collimated beam of the f/2 camera between the
  collimator and the camera.}
\end{figure*}

\begin{figure*}[ht]
\epsscale{0.7}
\plotone{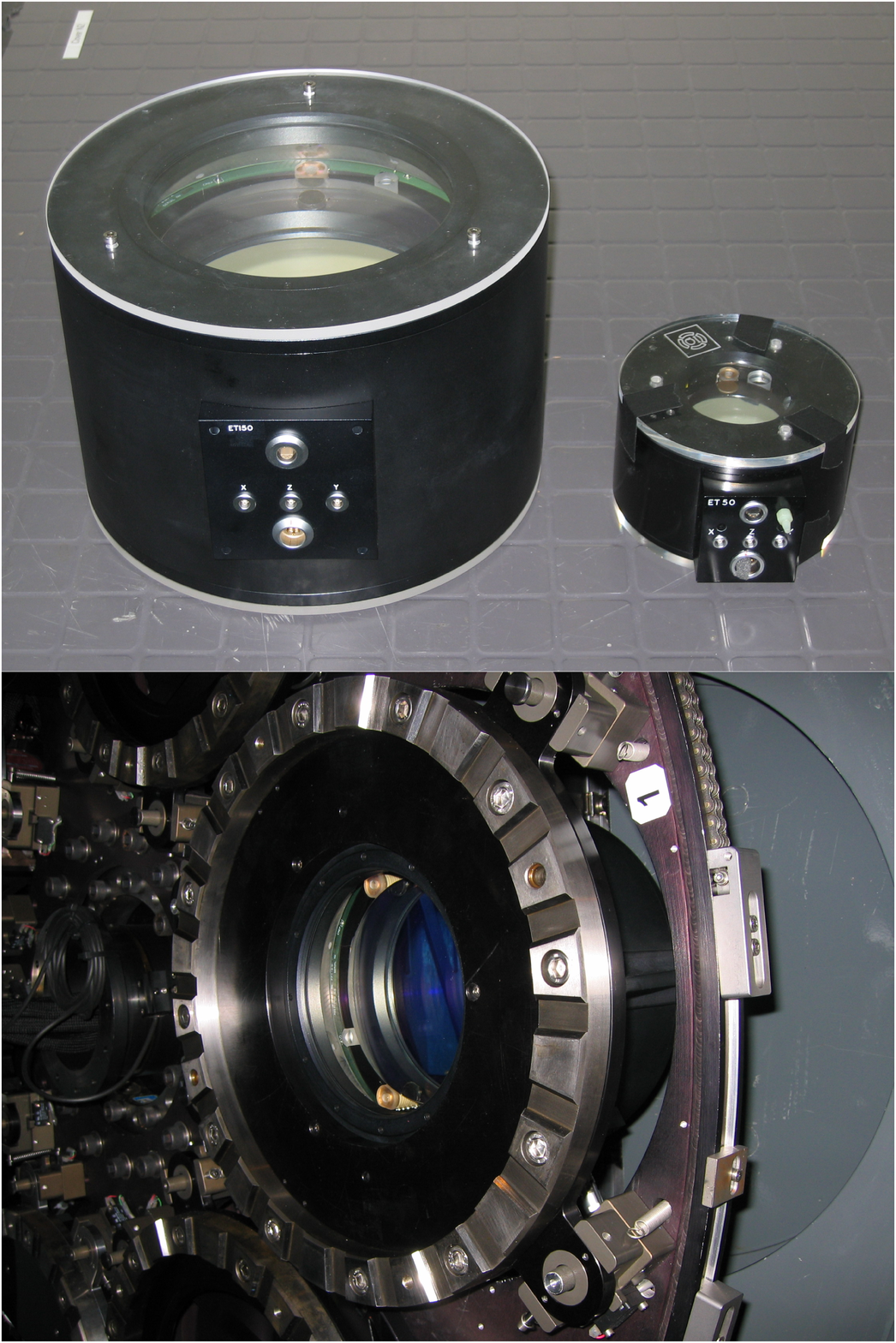}
\caption{({\em Top}) The ET150 etalon used for MMTF alongside the
  smaller ET50 etalon for comparison. ({\em Bottom}) The MMTF ET150
  etalon mounted in the IMACS disperser wheel.}
\end{figure*}

\begin{figure*}[ht]
\epsscale{0.8}
\plotone{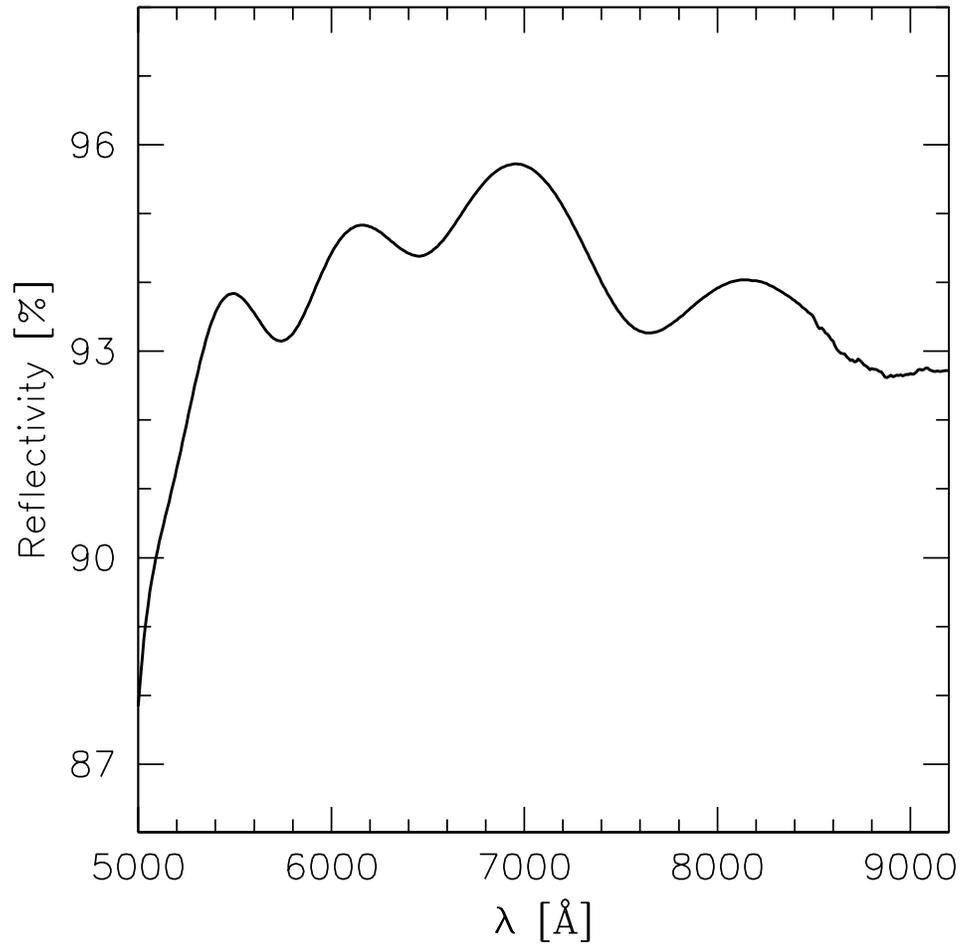}
\caption{The reflectivity curve of the MMTF ET150 etalon coating.}
\end{figure*}

\begin{figure*}[ht]
\epsscale{0.8}
\plotone{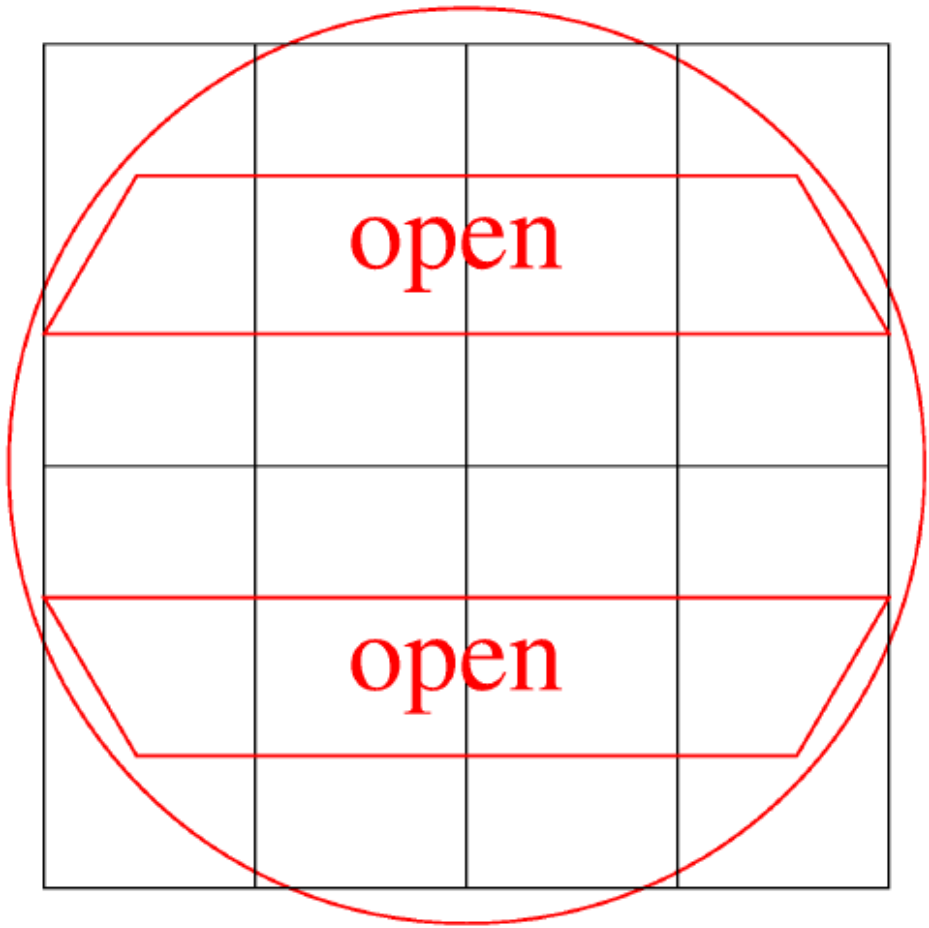}
\caption{Diagram of the aperture mask used in charge shuffling /
  frequency switching mode. The black outline shows the 8 CCD chips,
  and the red outline the aperture mask. The "open" segments are
  illuminated.}
\end{figure*}

\begin{figure*}[ht]
\epsscale{0.8}
\plotone{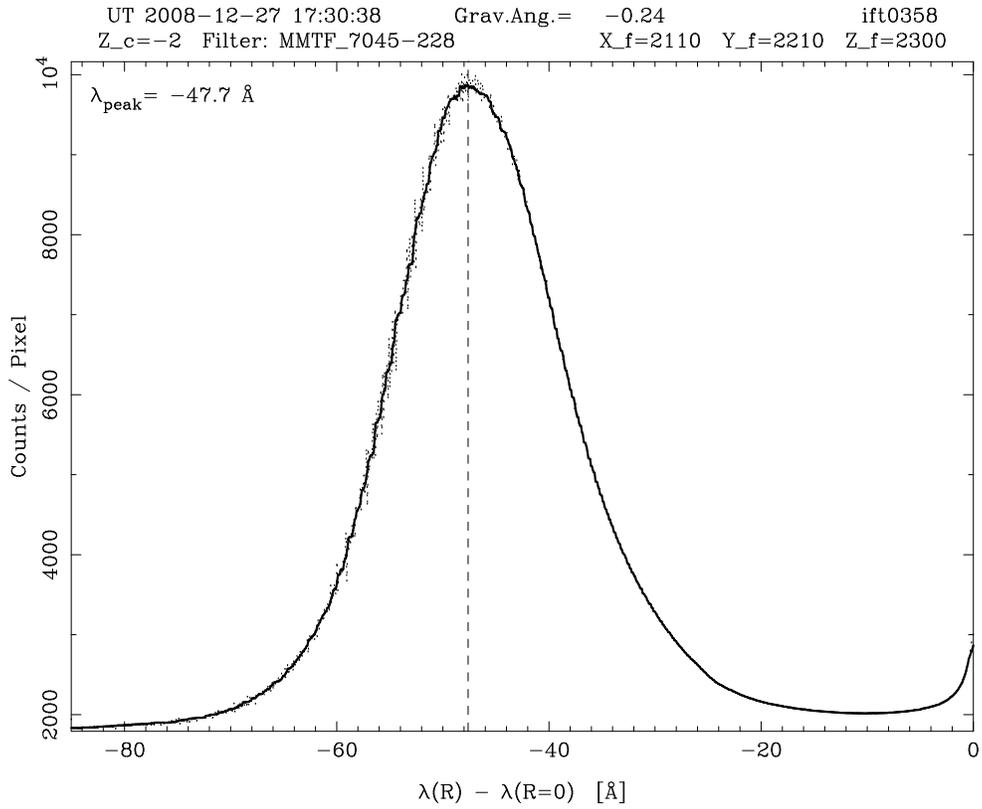}
\caption{Emission-line spectrum derived by azimuthally averaging
  reference rings from an arc lamp (Argon in this case).  The dotted
  line is the azimuthally averaged data, while the solid line is the
  data median smoothed by a 1 \AA\ window. Reference rings are taken
  at regular intervals to monitor wavelength drift and plate alignment
  during the night (see Figures 6 and 11).}
\end{figure*}

\begin{figure*}[ht]
\epsscale{0.8}
\plotone{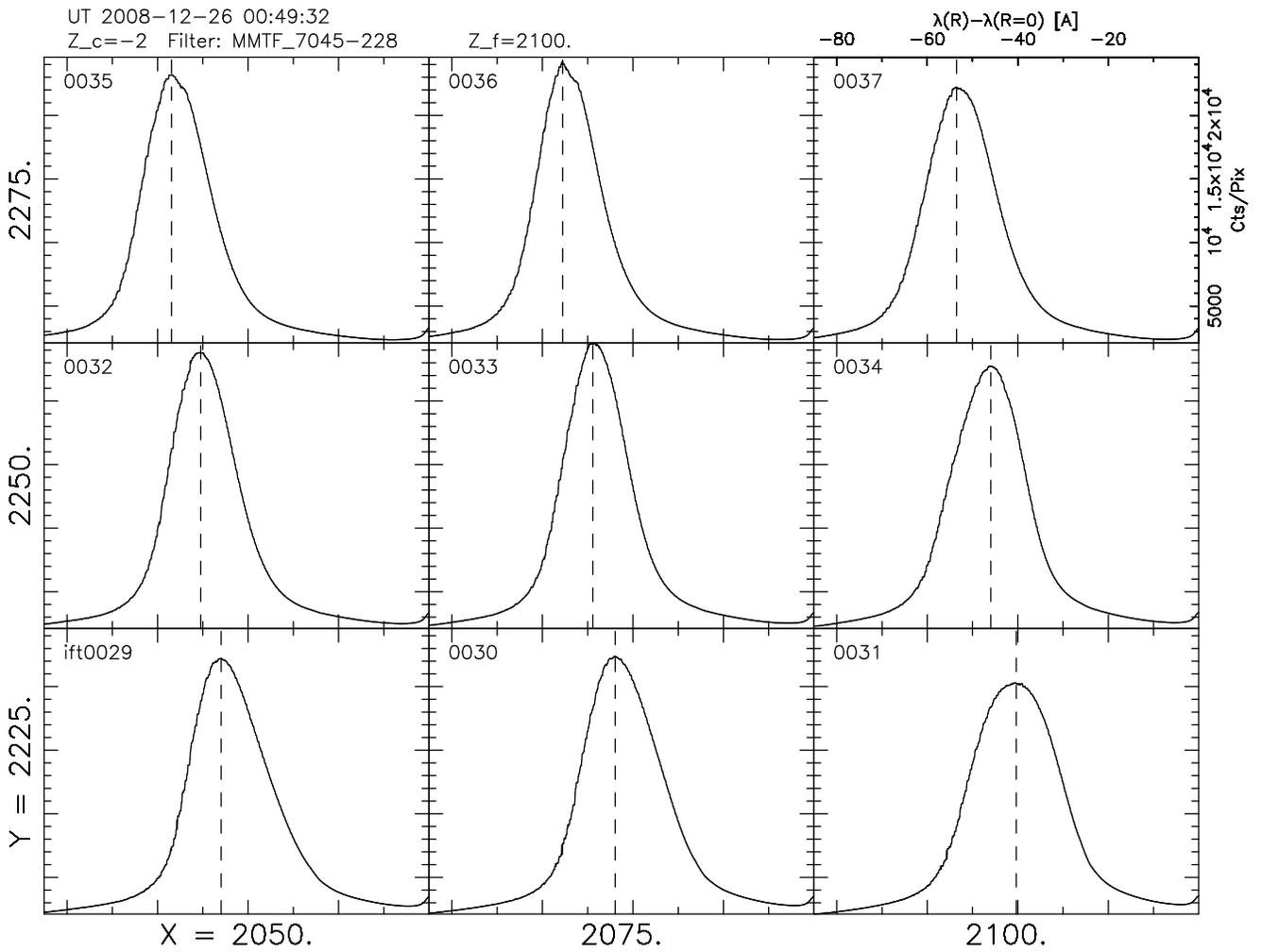}
\caption{Typical output from the plate alignment procedure. Reference
  rings are taken over a range in $X$ and $Y$. The profiles of
  emission features are compared by eye (and, where possible, with
  line fits) to find the narrowest and most symmetric profile (in this
  case, exposure \#033 in the center with $X \simeq$ 2075 and $Y
  \simeq$ 2250).}
\end{figure*}

\begin{figure*}[ht]
\epsscale{0.9}
\plotone{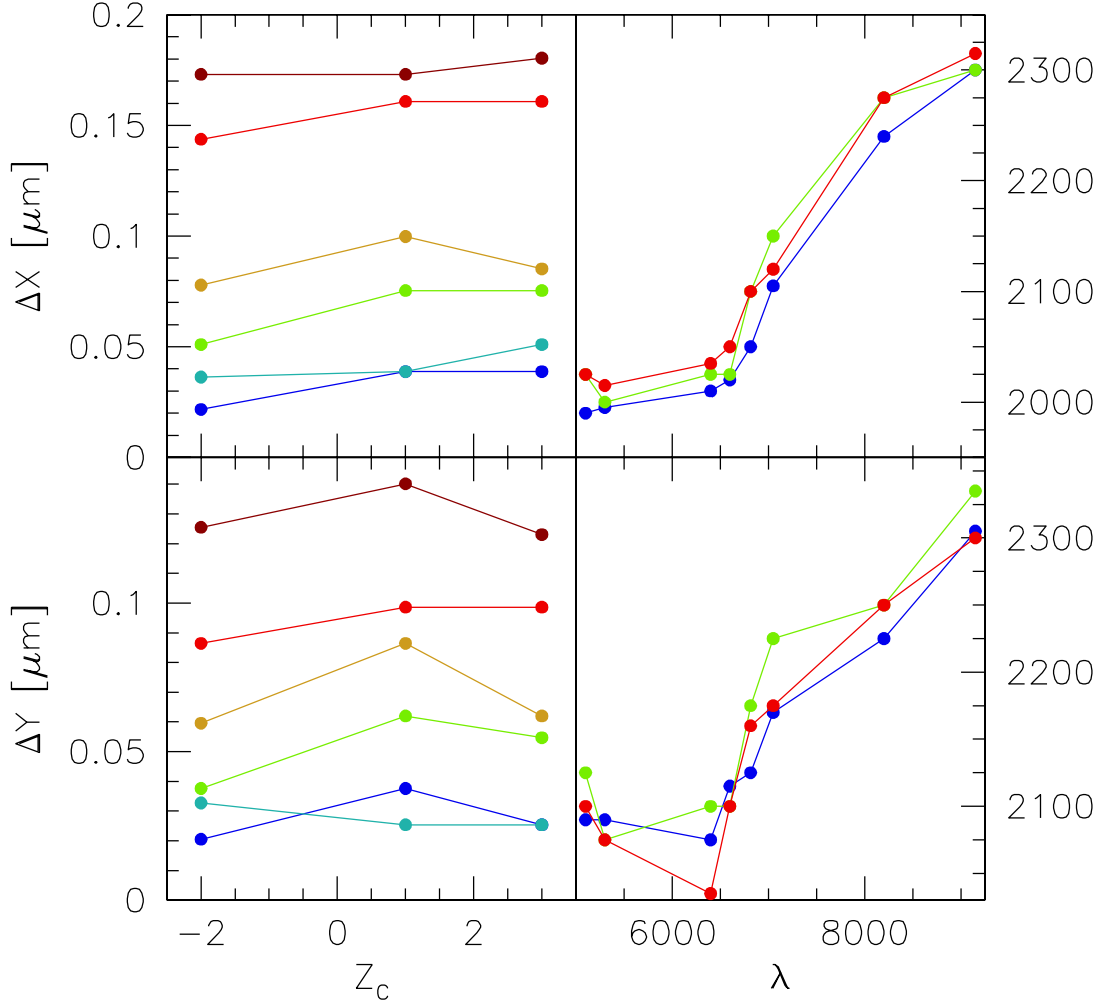}
\caption{Dependence of etalon alignment on etalon gap and wavelength,
  for a given etalon rotation angle. Shown on these plots are either
  the difference in $X$ and $Y$ positions in $\mu$m (on the left
  vertical axis) or in IMACS/MMTF computer control software units (on
  the right vertical axis). In the left panels, Z$_C$ is the coarse
  gap spacing, while the different colors represent the wavelength:
  5100 \AA\ (blue), 6400 \AA\ (turquoise), 6800 \AA\ (green), 7050
  \AA\ (orange), 8150 \AA\ (red) and 9150 \AA\ (deep red). In the
  right panels, the colors represent the coarse gap spacing: Z$_C$ =
  $-$2 (blue), +1 (green) and +3 (red). Note the change in X and Y of
  $\sim$0.10--0.15 $\mu$m for $\lambda \ga 6600$ \AA. This indicates
  that the relative angle between the effective reflecting surfaces of
  the etalon varies with wavelength.}
\end{figure*}

\begin{figure*}[ht]
\epsscale{0.8}
\plotone{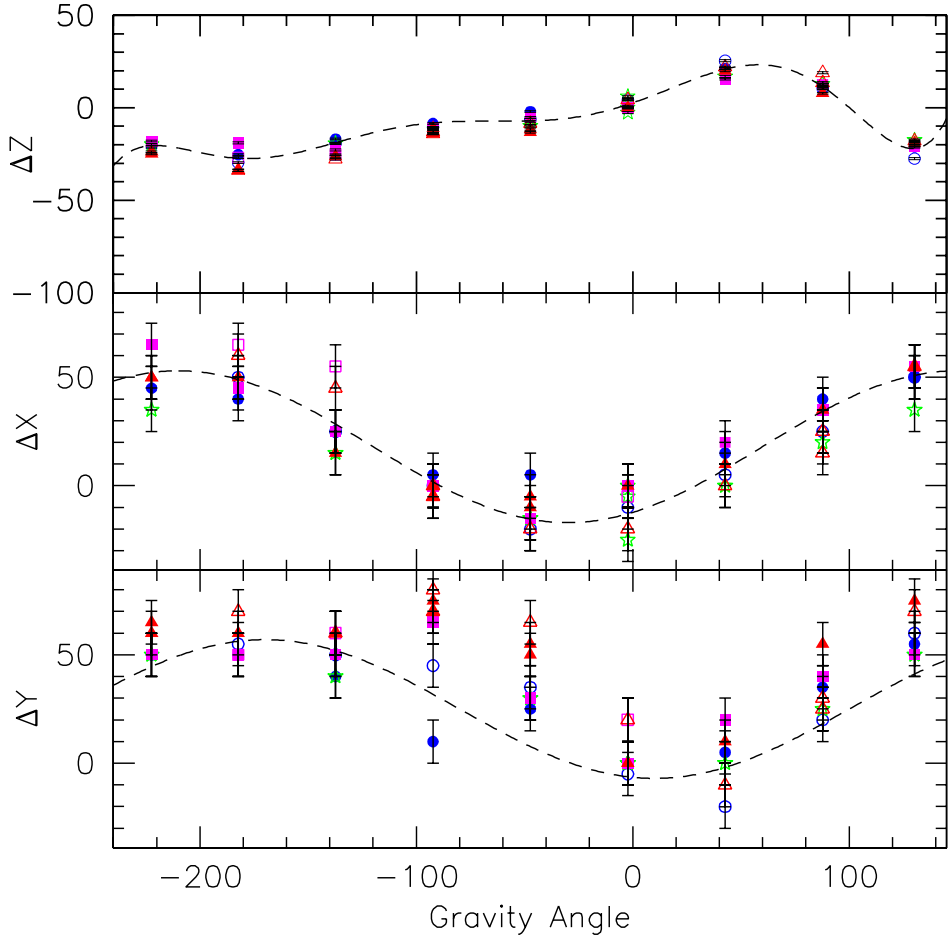}
\caption{Dependence of etalon alignment on gravity angle for a coarse
  gap spacing of Z$_c$ = $-$2.  The total amplitude of the change in
  alignment with gravity angle is $\sim$60 in IMACS computer control
  software units or $\sim$0.03 $\mu$m, but is generally highly
  reproducible (for reasons which remain unclear, the alignment values
  become difficult to predict for gravity angles between $-$120 and
  $-$20 degrees). The different point types and colors refer to
  sequences taken together.}
\end{figure*}

\begin{figure*}[ht]
\epsscale{0.8}
\plotone{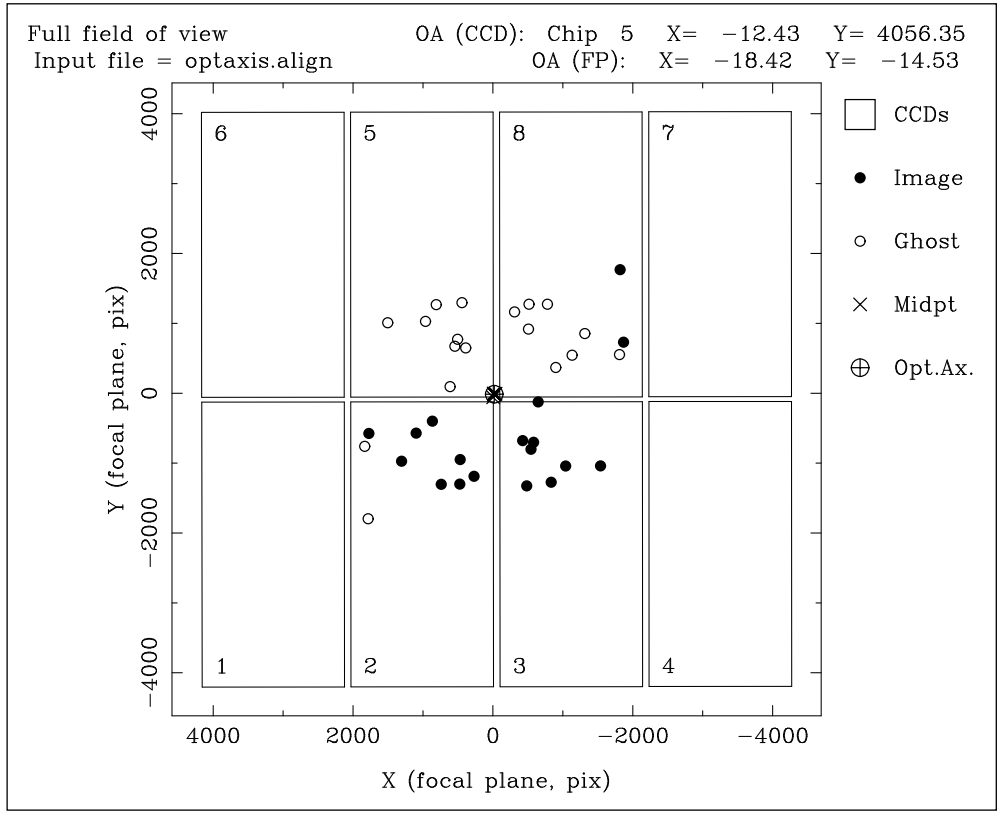}
\caption{Example of output from the procedure used to find the
  position of the optical axis. The position is typically accurate to
  $\pm$ 1 pixel (0$\farcs$2).}
\end{figure*}

\begin{figure*}[ht]
\epsscale{0.8}
\plotone{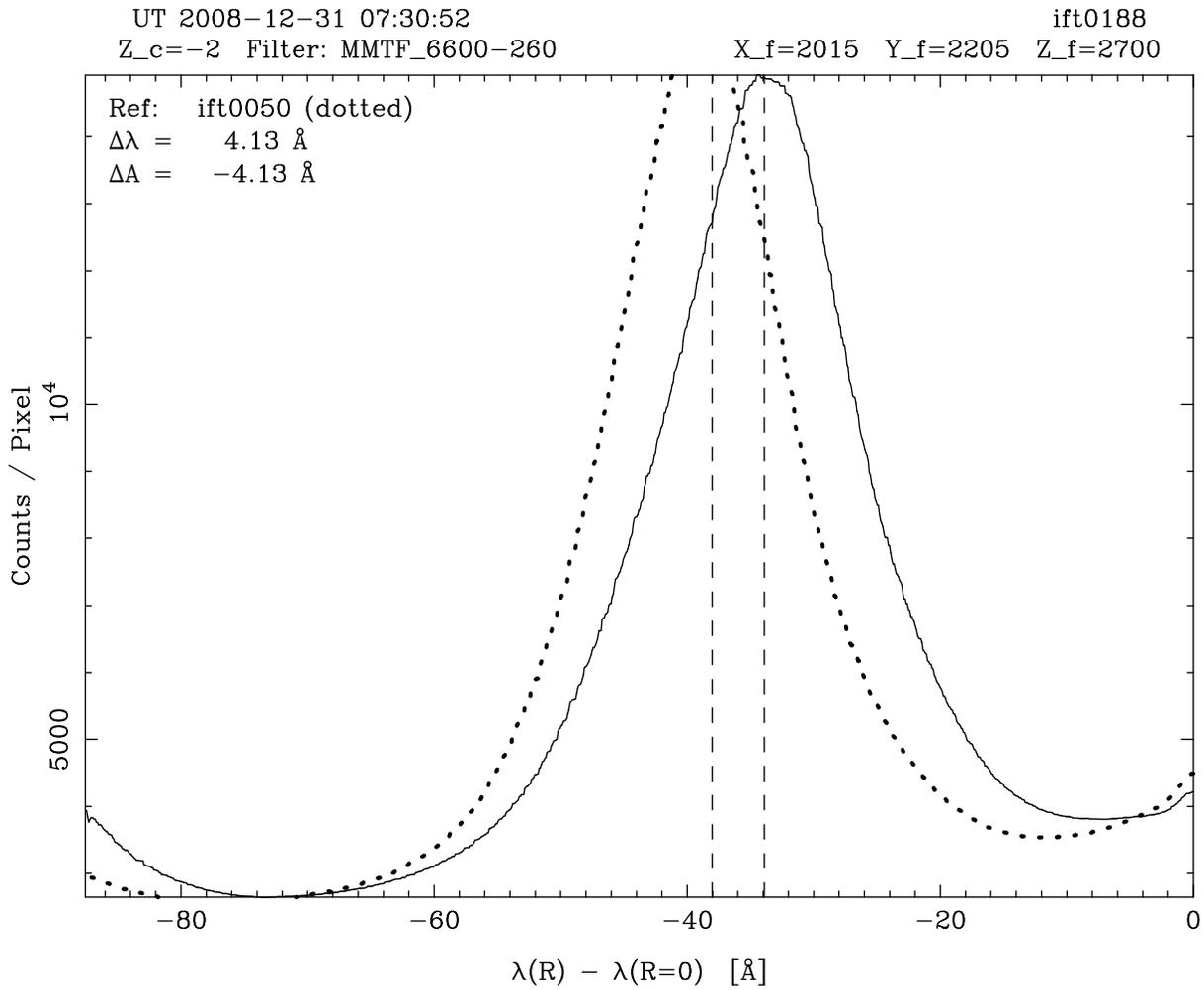}
\caption{Procedure to monitor wavelength drift. The emission-line
  profiles derived from reference rings taken at different times are
  compared and the shift is used to update the wavelength solution.}
\end{figure*}

\begin{figure*}[ht]
\epsscale{0.8}
\plotone{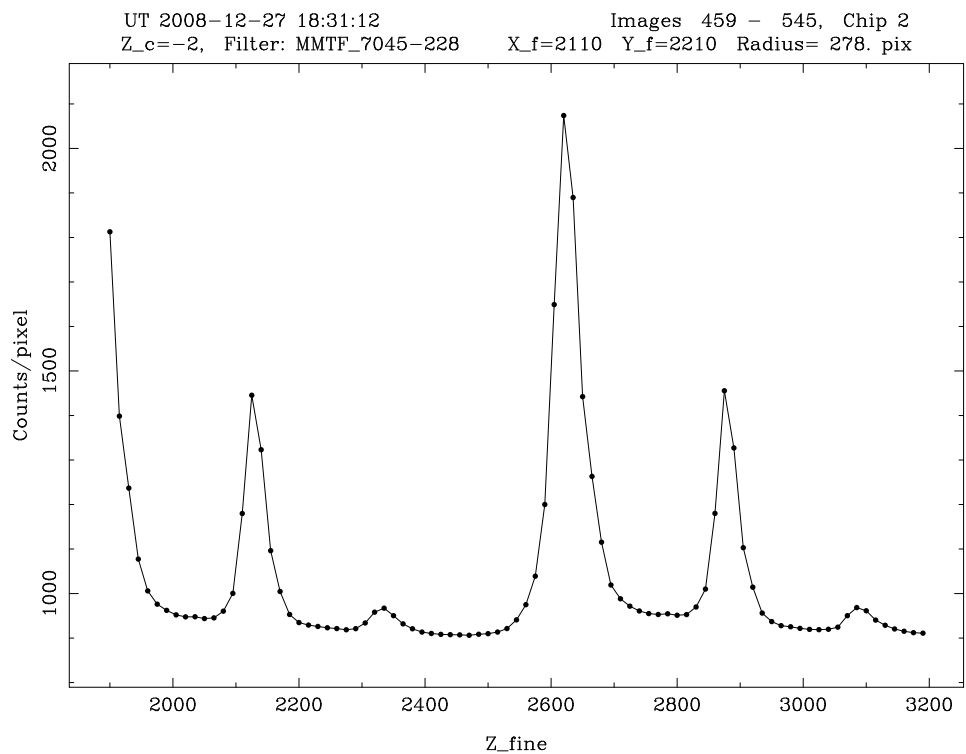}
\caption{Spectrum extracted from a ``data sausage'' of Argon. The
  data sausage is obtained by stepping the etalon spacing in small
  increments (Nyquist-sampled) through at least one FSR, but reading
  out only a small CCD subraster. The lines at $Z_{\rm fine} \sim$
  2120 and 2330 correspond to Ar~I $\lambda\lambda$7067.218, 7147.042,
  respectively. They are repeated at $Z_{\rm fine} \sim$ 2885 and
  3095. The line at $Z_{\rm fine} \sim$ 2615 is Ar~I
  $\lambda$6965.431. This particular sausage was obtained 278 pixels
  from the optical axis. The wavelength calibration procedure
  simultaneously acquires eight data sausages like this one at eight
  different positions (distances) from the optical axis and use them
  to derive the full wavelength solution.}
\end{figure*}

\begin{figure*}[ht]
\begin{minipage}{0.5\linewidth}
\includegraphics[width=1.0\linewidth]{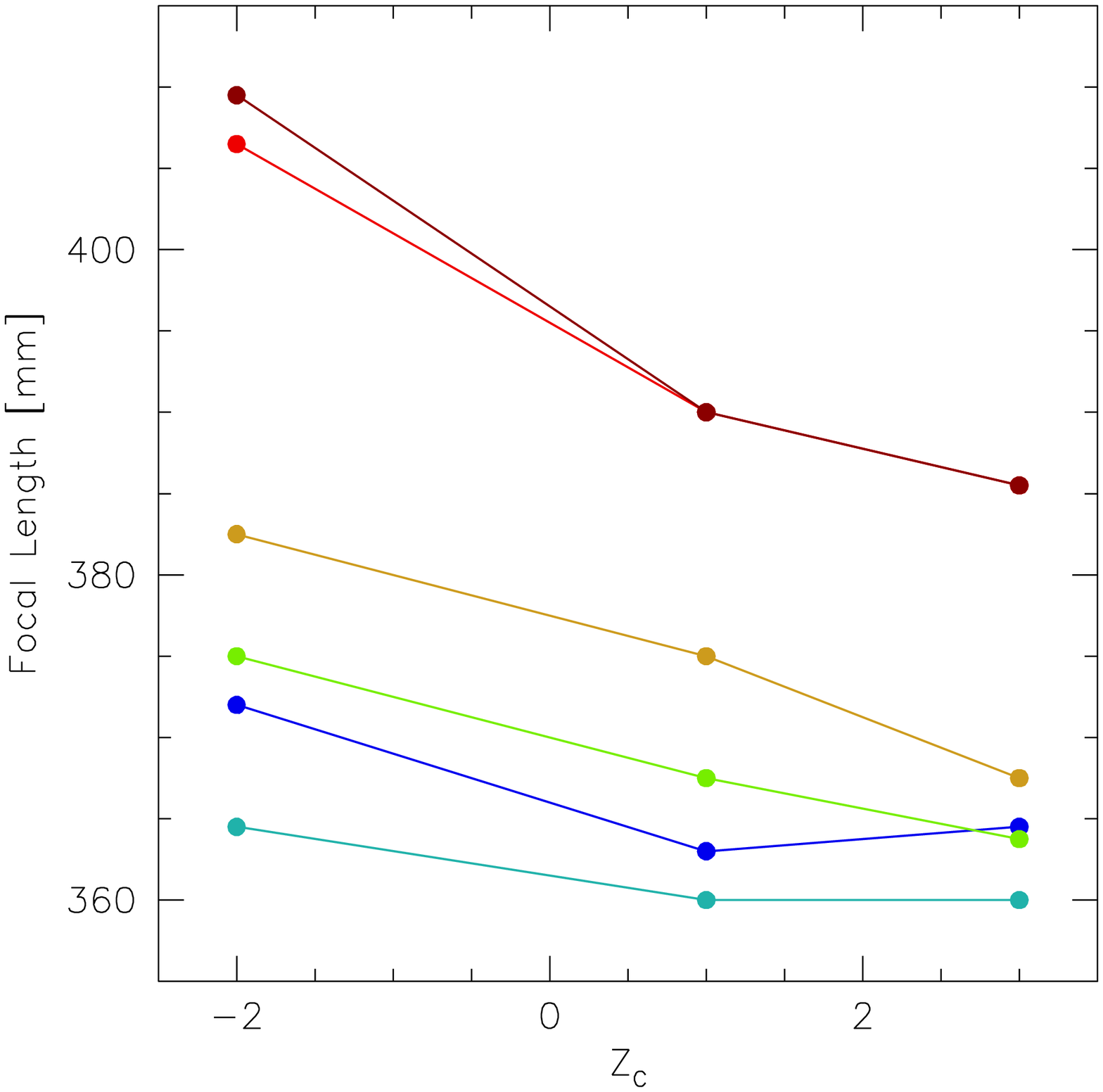}
\end{minipage}
\begin{minipage}{0.5\linewidth}
\includegraphics[width=1.0\linewidth]{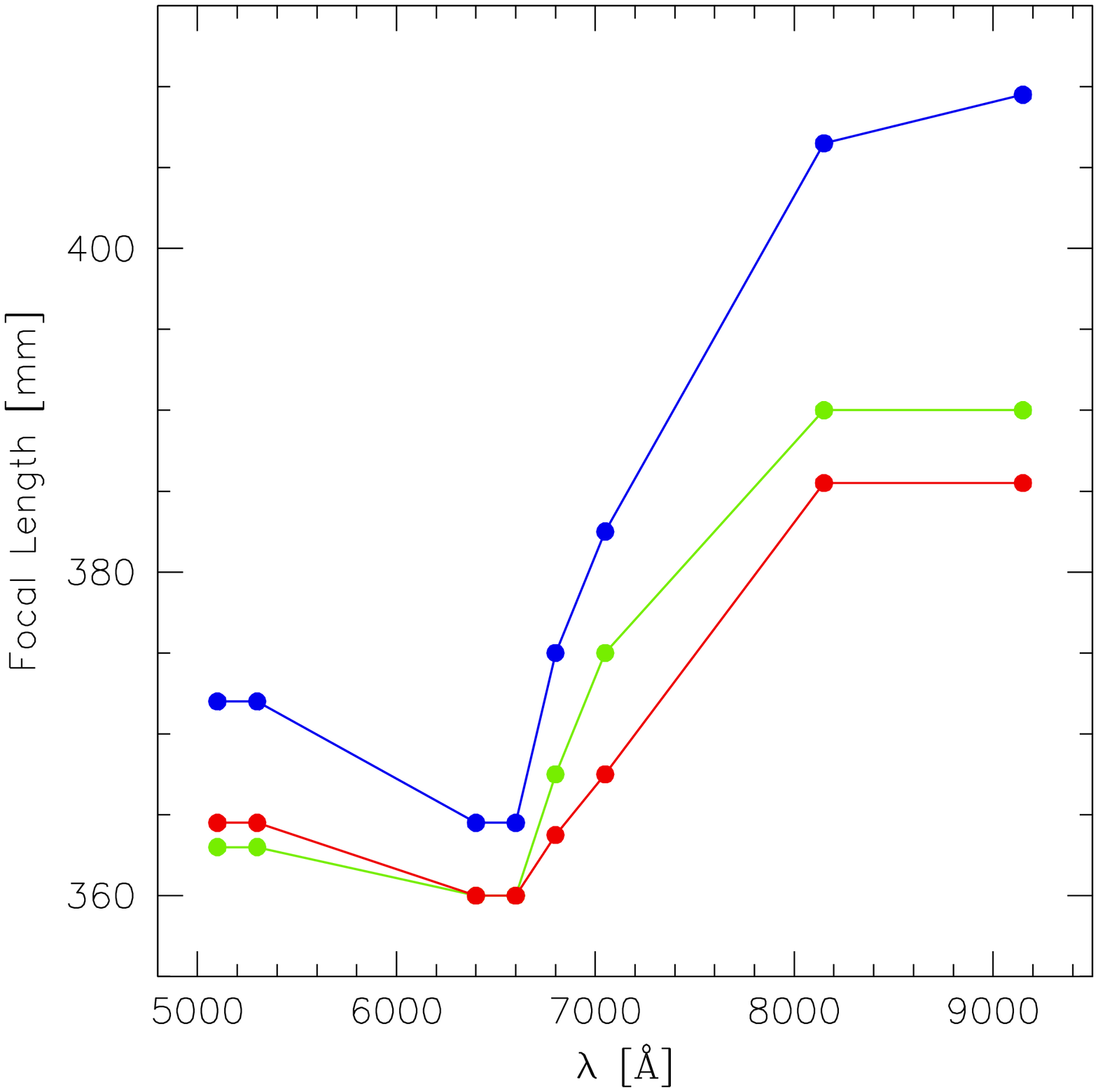}
\end{minipage}
\caption{Dependence of the focal length (in mm) on etalon gap and
  wavelength. In the left panel, Z$_C$ is the coarse gap spacing,
  while the different colors represent the wavelength: 5100 \AA\
  (blue), 6400 \AA\ (turquoise), 6800 \AA\ (green), 7050 \AA\
  (orange), 8150 \AA\ (red) and 9150 \AA\ (deep red). In the right
  panel, the colors refer to the coarse gap spacing: Z$_C$ = $-$2
  (blue), +1 (green) and +3 (red). These results suggest slight
  chromatic aberrations due in part to MMTF. }
\end{figure*}


\clearpage

\begin{references}

\refpar
Atherton, P.D.\ et al.\ 1981, Opt.\ Eng., 20, 806
\refpar
Bland, J., \& Tully, R. B. 1989, AJ, 98, 723
\refpar
Bland-Hawthorn, J. 2000a, in Imaging the Universe in Three
Dimensions. Proc.\ ASP Conf. Vol. 195. Edited by W. van
Breugel and J. Bland-Hawthorn, p. 34
\refpar
Bland-Hawthorn, J. 2000b, in Encyclopedia of Astronomy \&
Astrophysics, MacMillan and Institute of Physics Publishing.
\refpar
Bland-Hawthorn, J., \& Jones, D. H. 1998a, PASA, 15, 44
\refpar
Bland-Hawthorn, J., \& Jones, D. H. 1998b, SPIE, 3355, 855
\refpar
Bland-Hawthorn, J., \& Kedziora-Chudczer, L. 2003, PASA, 20, 242
\refpar
Cepa, J., et al. 2007, RMxAC, 29, 168
\refpar
Dopita, M. A., \& Hua, C. T. 1997, ApJS, 108, 515
\refpar
Dressler, A., Hare, T., Bigelow, B. C., \& Osip, D. J. 2006, SPIE, 6269, 62690F 
\refpar
Hamuy, M. et al.\ 1992, PASP, 104, 533
\refpar
Hamuy, M. et al.\ 1994, PASP, 106, 566
\refpar
Jacquinot, P. 1954, J Opt Sci Amer, 44, 761
\refpar
Jones, D. H.,  \& Bland-Hawthorn, J. 1998, PASP, 110, 1059
\refpar
Jones, D. H., Shopbell, P. L., \& Bland-Hawthorn, J. 2002, MNRAS, 329,
759
\refpar
Macleod, H. A.\ 2001, ``Thin-Film Optical Filters'', CRC Press, 3rd edition, 641 p.
\refpar
Oke, J. B.\ 1990, AJ, 99, 1621
\refpar
Rangwala, N., Williams, T. B., Pietraszewski, C., \& Joseph,
C. L. 2008, AJ, 135, 1825
\end{references}
\end{document}